\documentclass[%
 aip,
 amsmath,amssymb,
 reprint,%
]{revtex4-1}

\usepackage{graphicx}% Include figure files
\usepackage{dcolumn}% Align table columns on decimal point
\usepackage{bm}% bold math
\usepackage{physics}

\usepackage[utf8]{inputenc}
\usepackage[T1]{fontenc}
\usepackage{mathptmx}
\usepackage{amsmath}
\usepackage{etoolbox}
\usepackage{xcolor}
\usepackage{ulem}
\usepackage{siunitx}

\newcommand{\rr}{\boldsymbol{R}}
\newcommand{\pp}{\boldsymbol{P}}
\newcommand{\hh}{\hat{H}}
\newcommand{\D}{\boldsymbol{d}}

\makeatletter
\def\@email#1#2{%
 \endgroup
 \patchcmd{\titleblock@produce}
  {\frontmatter@RRAPformat}
  {\frontmatter@RRAPformat{\produce@RRAP{*#1\href{mailto:#2}{#2}}}\frontmatter@RRAPformat}
  {}{}
}%
\makeatother
\begin{document}

\preprint{AIP/123-QED}

\title{Interband and intraband transitions, as well as charge mobility in driven two-band model with electron phonon coupling}
% Force line breaks with \\
\author{Yu Wang}
%\email{wangyu19@westlake.edu.cn}
 \affiliation{Department of Chemistry, School of Science, Westlake University, Hangzhou 310024 Zhejiang, China}%Lines break automatically or can be forced with \\
 \affiliation{Institute of Natural Sciences, Westlake Institute for Advanced Study, Hangzhou 310024 Zhejiang, China}
\author{Wenjie Dou}%
 \email{douwenjie@westlake.edu.cn}
 \affiliation{Department of Chemistry, School of Science, Westlake University, Hangzhou 310024 Zhejiang, China}
 \affiliation{Institute of Natural Sciences, Westlake Institute for Advanced Study, Hangzhou 310024 Zhejiang, China}
 \affiliation{Department of Physics, School of Science, Westlake University, Hangzhou 310024 Zhejiang, China}

\date{\today}% It is always \today, today,

\begin{abstract}
Interband and intraband transitions are fundamental concepts in the study of electronic properties of materials, particularly semiconductors and nanomaterials. These transitions involve the movement of electrons between distinct energy states or bands within a material.
Besides, charge mobility is also a critical parameter in materials science and electronics.
A thorough understanding of these transitions and mobility is critical for the development and optimization of advanced electronic and optoelectronic devices.
In this study, we investigate the influence of external periodic drivings on interband and intraband transitions, as well as charge mobility, within a driven two-band model that includes electron-phonon coupling. These external periodic drivings can include a periodic laser field, a time-varying magnetic or electric field, or an alternating current (AC) voltage source.
We have developed the Floquet surface hopping (FSH) and Floquet mean field (FMF) methods to simulate electronic dynamics under various drivings in both real and reciprocal spaces. Our findings demonstrate that periodic drivings can enhance interband transitions while suppressing intraband transitions. Additionally, charge mobility is restrained by these external periodic drivings in the driven two-band model.

\end{abstract}

\maketitle

\section{\label{sec:level1}Introduction}
Carrier dynamics, which refers to the behavior and motion of charge carriers (such as electrons and holes) in solid materials, plays a crucial role in determining the electronic, optical, and transport properties of materials\cite{ulbricht2011carrier,liao2020understanding,zhu2019strain}.
For solid-state materials, electron-phonon couplings significantly influence the electronic and thermal characteristics\cite{quan2021impact,giustino2017electron}. Phonon-assisted transitions involve the absorption or emission of a phonon to conserve momentum, facilitating the recombination of electrons and holes. This process is critical in silicon-based devices, such as transistors and solar cells\cite{ghosh2017free,wright2016electron}.
In this study, we mainly focus on phonon-assisted interband and intraband transition\cite{wehrenberg2002interband}, as well as charge mobility\cite{wang2010computational} under the influence of external periodic driving. External periodic driving, such as an oscillating electric field, an AC voltage source, a periodic laser field, etc., can influence a wide range of systems, leading to diverse and often complex behaviors\cite{creffield2007quantum,bordia2017periodically}. Understanding these responses is crucial in designing and controlling systems.

Interband transitions, involving the excitation of charge carriers from the valence band to the conduction band, are fundamental to the functionality of optoelectronic devices such as light-emitting diodes (LEDs)\cite{kim2018improved}, laser diodes\cite{meyer2020interband}, and photovoltaic cells\cite{wu2012strong}. In particular, the performance of a solar cell is critically dependent on the efficiency with which it utilizes interband transitions to convert incident photons into electrical energy\cite{okada2015intermediate}. Upon absorption of light, excitons are generated; these bound electron-hole pairs must then be efficiently dissociated into free charge carriers (electrons and holes) to facilitate the generation of an electric current\cite{grancini2013hot}.
In a periodically driven lattice, such as a semiconductor crystal subjected to an oscillating electric field, electrons can experience Floquet-Bloch oscillations\cite{gomez2013floquet}. This can be used to control electronic properties of materials in ultrafast electronics and optoelectronics\cite{wang2013observation,upreti2020topological,lucchini2022controlling}.

Intraband transitions refer to the movement of charge carriers (electrons or holes) within the same energy band, typically within the conduction band or the valence band. These transitions enable a variety of applications, especially in the infrared and terahertz regions. They provide essential functionality in sensing, imaging, communication, and advanced electronic devices.
With periodic driving, the electron's energy and momentum can be modulated, leading to dynamic phenomena such as high harmonic generation (HHG)\cite{nakagawa2022size,hirori2024high} and Bloch oscillations\cite{leo1998interband}. For example, Ghimire et al. have demonstrated HHG in bulk crystals driven by intense laser fields, showing that electrons driven within the same band can produce high harmonics of the driving frequency\cite{ghimire2011observation}.

Charge mobility is a critical parameter in the performance of semiconductor devices, including transistors, solar cells, and LEDs\cite{shuai2014charge}.
Periodic optical driving, such as intense light or laser pulses, can excite charge carriers to higher energy states, affecting their mobility. This can lead to phenomena like the Franz-Keldysh effect, where the absorption edge of a semiconductor shifts under the influence of an electric field, altering carrier dynamics\cite{jauho1996dynamical,nordstrom1998excitonic}.
Under certain conditions, periodic driving can lead to coherent transport where charge carriers experience fewer scattering events, thereby increasing mobility\cite{potanina2021thermodynamic}.
Periodic optical driving can induce photoconductivity, where the mobility of photo-generated carriers is modulated by the periodic light, enhancing the performance of photodetectors and photovoltaic device\cite{luo2019high}.

To address time periodic system, we can use Floquet theory, which is a powerful mathematical framework used to analyze the behavior of periodic systems in various fields, including physics, engineering, and mathematics\cite{oka2019floquet,klausmeier2008floquet}.
For example, Leskes et al. introduce Floquet theory as a means to enhance the understanding and implementation of solid-state nuclear magnetic resonance (NMR) spectroscopy\cite{leskes2010floquet}.
Shu et al. present an experimental study on the observation of Floquet Raman transitions in a solid-state spin system subjected to periodic driving, which demonstrate the potential of Floquet engineering for controlling quantum states\cite{shu2018observation}.
In the work of Rechtsman et al., Floquet theory offers a powerful tool to engineer and control topological phases in photonic systems\cite{rechtsman2013photonic}.
Tiwari et al. employed Floquet theory to study the optical absorption of laser-dressed solids\cite{tiwari2023floquet}.

In this study, within the frame of Floquet theory, we developed Floquet surface hopping (FSH) and Floquet mean field (FMF) methods to simulate interband and intraband transitions, as well as charge mobility under the influence of periodic driving.
Both of these two methods can be employed in real and reciprocal space.
For simulations of interband and intraband transitions, conducting them in reciprocal space is more convenient. However, for simulations of charge mobility, conducting them in real space is more convenient.
The surface hopping algorithm has previously been successfully applied in solid-state dynamics\cite{zheng2023multiple,xie2022surface,krotz2022reciprocal,chen2024floquet}. Although the mean field method has been used in solid-state dynamics calculations\cite{krotz2021reciprocal,chen2024floquet}, it has limitations in certain parameter regimes.
Herein, we extend both surface hopping and mean field algorithms to two-band model in Floquet space.
External periodic driving helps us control complex systems beyond their usual stable states. Our two methods are important for understanding these dynamic effects. Also, these methods can improve our basic knowledge of the systems and can help optimize their performance in various technological applications.

\section{Theory}
\subsection{Model Hamiltonian}
In this study, we focus on the one-dimensional lattice model, with two sites per unit cell, $J$ and $B$.
The total Hamiltonian with electron-phonon coupling and time periodic driving is given as
\begin{equation}
    \hh_{tot}(t) = \hh_{el}(t) + \hh_{el-ph}(\rr) + H_{ph}(\rr,\pp).
\end{equation}
Here, $\hh_{el}(t)$ is the electronic Hamiltonian, which is subject to a time periodic driving. Consequently, we have $\hh_{el}(t)=\hh_{el}(t+T)$, where $T$ is the period of the driving. $\hh_{el-ph}(\rr)$ is the electron-phonon coupling, which is dependent on classical nuclear position $\rr$. In this model, we consider dispersionless Einstein phonons $H_{ph}(\rr,\pp)$, which is a simplified theoretical model where all phonons in the crystal lattice have the same frequency (as in the Einstein model) and there is no dispersion (frequency does not depend on wavevector).
The phonon Hamiltonian is depend on both nuclear position $\rr$ and momentum $\pp$.

In real space, the electronic Hamiltonian $H_{el}(t)$ is formulated as,
\begin{equation}
\begin{split}
    \hh_{el}(t) = & -\sum_n(J_1+J\cos(\Omega t))\ket{n,B}\bra{n,A} + \\ & \sum_n(J_2-J\cos(\Omega t))\ket{n+1,A}\bra{n,B}+h.c.,
\end{split}
\end{equation}
here, $J_1$ is the hopping energy inside the cell, and $J_2$ is the energy required to hop between unit cells.
$J\cos(\omega t)$ is the time periodic driving with strength $J$ and frequency $\Omega$.

The electron-phonon coupling, which is dependent on classical nuclear positions $\rr=(r_1,r_2,...,r_n)$, is expressed as,
\begin{equation}
\begin{split}
    \hh_{el-ph} =  \sum_n g\sqrt{2\omega^2} ( & r_{n,A}\ket{n,A}\bra{n,A} 
    + \\ &r_{n,B}\ket{n,B}\bra{n,B}),
\end{split}
\end{equation}
here, we consider the local electron-phonon coupling. $g$ is the dimensionless coupling parameter, $\omega$ is the nuclear oscillation frequency. The vibrational reorganization energy is $g^2\omega$.
For phonons, we consider the classical, noninteracting, harmonic oscillations,
\begin{equation}
    H_{ph} = \sum_n\frac{1}{2}(p_{n,A}^2+\omega^2r_{n,A}^2+p_{n,B}^2+\omega^2r_{n,B}^2).
\end{equation}

By employing Fourier transformation, we can transfer the system from real space to reciprocal space,
\begin{equation}
    f(k) = \frac{1}{\sqrt{N}}\sum_{n=1}^N e^{ikr_n}f(r).
\end{equation}
Reciprocal space simplifies the treatment of periodic systems, such as crystals, due to the inherent periodicity in the structure.
According to Ref. \citenum{krotz2022reciprocal}, in reciprocal space, the nuclear position $r_k$ and momentum $p_k$ have the following relationships with their counterparts in real space,
\begin{equation}\label{rk}
    r_k =\frac{1}{\sqrt{N}}\sum_{n=1}^{N} \left(r_n \cos(kn)-\frac{p_n}{\omega}\sin(kn)\right),
\end{equation}
\begin{equation}\label{pk}
    p_k =\frac{\omega}{\sqrt{N}}\sum_{n=1}^N \left(\frac{p_n}{\omega}\cos(kn)+r_n\sin(kn)\right),
\end{equation}
where $k$ is the wavevector ranging from $-\pi/a$ to $\pi/a$, $J$ is the lattice constant. 

The electron part in reciprocal space becomes,
\begin{equation}
\begin{split}
    \hh_{el} =& -\sum_k((J_1+J\cos(\Omega t))+ \\ & e^{ik}(J_2-J\cos(\Omega t)))\ket{k,A}\bra{k,B}+ h.c.),
\end{split}
\end{equation}
the electron-phonon coupling becomes,
\begin{equation}\label{el-ph-k}
\begin{split}
    \hh_{el-ph} = & \frac{g\sqrt{\omega}}{\sqrt{2N}}\sum_{k,k'}\ket{k+k',A}\bra{k,A}\times \\& (\omega(r_{-k',A}+r_{k',A})-i(p_{-k',A}-p_{k',A})) + \\& 
    \frac{g\sqrt{\omega}}{\sqrt{2N}}\sum_{k,k'}\ket{k+k',B}\bra{k,B}\times \\& (\omega(r_{-k',B}+r_{k',B})-i(p_{-k',B}-p_{k',B})),
\end{split}
\end{equation}
and the phonon part becomes,
\begin{equation}
    H_{ph} = \sum_k\frac{1}{2}(p_{k,A}^2+\omega^2q_{k,A}^2+p_{k,B}^2+\omega^2q_{k,B}^2).
\end{equation}

\subsection{Floquet theory}

Floquet theory is a crucial mathematical framework for analyzing differential equations with periodic coefficients. For systems subject to periodic driving forces, where the coefficients of the governing differential equations are periodic functions of time, Floquet theory is essential for understanding their behavior.
Here, we follow Ref. \citenum{mosallanejad2023floquet} to construct Floquet Hamiltonian $H_{tot}^F$.
Firstly, we need introduce two main operators, Fourier number operators $\hat{N}$ and the Fourier ladder operators $\hat{L}_u$. They have following properties,
\begin{equation}
    \hat{N}\ket{u}=u\ket{u}, \hat{L}_u\ket{v}=\ket{u+v},
\end{equation}
where $\ket{u}$ is the basis set in the Fourier space.
Then the Hamiltonian and density operator in Floquet representation would be
\begin{equation}
    \hh^F = \sum_u\hh^{(u)}\hat{L}_u + \hat{N}\hbar\Omega,
\end{equation}
\begin{equation}
    \hat{\rho}^F(t) = \sum_u\hat{\rho}^{(u)}(t)\hat{L}_u.
\end{equation}
Here, $\hh^{(u)}$ and $\hat{\rho}^{(u)}(t)$ are the Fourier expansion coefficients in $\hh(t)=\sum_u\hh^{(u)}e^{iu\omega t}$ and $\hat{\rho}(t)=\sum_u\hat{\rho}^{(u)}(t)e^{iu\omega t}$.

In the Floquet representation, the equation of motion (EOM) of $\hat{\rho}^F(t)$ has the same form with $\hat{\rho}(t)$, which follows the Liouville-von Neumann (LvN) equation:
\begin{equation}\label{F-EOM}
    \frac{\partial}{\partial t}\hat{\rho}^F(t) = -\frac{i}{\hbar}[\hh^F,\hat{\rho}^F(t)].
\end{equation}
Instead of Schr\"{o}dinger equation, we evolve density matrix according to the LvN equation, which can describe both pure and mixed states via the density matrix. In previous study, we have successfully employ Floquet LvN equation to study nonadiabatic dynamics near metal surface with periodic drivings\cite{wang2023nonadiabatic,wang2024nonadiabatic}.

\subsection{Floquet mean field (FMF)}

In real space, classical particles move in the average potential created by the quantum particles in
Floquet mean field (FMF) dynamics. We can express FMF in real space as follows,
\begin{eqnarray}
\begin{aligned}
    &\dot{r}_n = p_n, \\
    & \dot{p}_n = -Tr(\Delta_{r_n}\hh^F\hat{\rho}^F) - \omega^2r_n.
\end{aligned}    
\end{eqnarray}
Here, $\hh^F$ including pure electron and electron-phonon coupling parts.

In reciprocal space, we note that $\hh_{el-ph}$ is related not only to $r_k$, but also to $p_k$, as evident from Eq. \ref{el-ph-k}. Therefore, the EOM in FMF dynamics becomes,
\begin{eqnarray}
\begin{aligned}
    &\dot{r}_k = Tr(\Delta_{p_k}\hh^F\hat{\rho}^F) + p_k, \\
    & \dot{p}_k = -Tr(\Delta_{r_k}\hh^F\hat{\rho}^F) - \omega^2r_k.
\end{aligned}    
\end{eqnarray}

\subsection{Floquet surface hopping (FSH)}

Unlike the FMF method, the Floquet surface hopping (FSH) method evolves classical particles on adiabatic surfaces with hopping rates between these surfaces.
In real space, the hopping rate from surface $\alpha$ to surface $\beta$ can be determined as,
\begin{equation}
    k_{\alpha\rightarrow \beta} = \Theta\left(-2\Re \left(\pp\cdot \D_{\alpha\beta}\frac{\hat{\rho}^{{F(ad)}}_{\beta\alpha}}{\hat{\rho}^{{F(ad)}}_{\alpha\alpha}}\right)\right),
\end{equation}
where $\Theta$ is the Heaviside function:
\begin{equation}
    \Theta(x) = 
    \begin{cases}
        x & \text{x>0} \\
        0 & \text{x<0},
    \end{cases}
\end{equation}
The $\pp\cdot \D_{\alpha\beta}$ comes from $\bra{\Psi_{\alpha}^{F(ad)}}\frac{\partial}{\partial t}\ket{\Psi_{\beta}^{F(ad)}}$, where $\Psi_{\alpha}^{F(ad)}$ is the adiabatic Floquet wavefunction which is dependent on $r_n$.
We have $\hat{\rho}^{{F(ad)}}_{\beta\alpha}=\ket{\Psi_{\beta}^{F(ad)}}\bra{\Psi_{\alpha}^{F(ad)}}$.
The derivative coupling $\D_{\alpha\beta}$ has the form as,
\begin{equation}
    \D_{\alpha\beta} = \bra{\Psi_{\alpha}^{F(ad)}}\frac{\partial}{\partial r_n}\ket{\Psi_{\beta}^{F(ad)}},
\end{equation}
%\begin{equation}
%    \D_{\alpha\beta} = \frac{-\bra{\Psi_{\alpha}^{F(ad)}}\frac{\partial \hat{H}^{F}}{\partial r_n}\ket{\Psi_{\beta}^{F(ad)}}}{E_{\alpha}-E_{\beta}},
%\end{equation}
Following Tully's surface hopping algorithm, after a hopping event, the nuclear momentum is rescaled in the direction of $\D_{\alpha\beta}$,
\begin{equation}
    p^{new} = p - \kappa \D_{\alpha\beta}/|\D_{\alpha\beta}|.
\end{equation}

In reciprocal space, the classical position and momentum become scrambled as shown in Eqs. \ref{rk} and \ref{pk}. Such that $\Psi_{\alpha}^{F(ad)}$ is dependent on both $r_k$ and $p_k$.
Consequently,
$\bra{\Psi_{\alpha}^{F(ad)}}\frac{\partial}{\partial t}\ket{\Psi_{\beta}^{F(ad)}}$ becomes
\begin{equation}
\begin{split}
     \frac{\partial r_k}{\partial t}\bra{\Psi_{\alpha}}\frac{\partial}{\partial r_k}\ket{\Psi_{\beta}} +\frac{\partial p_k}{\partial t}\bra{\Psi_{\alpha}}\frac{\partial}{\partial p_k}\ket{\Psi_{\beta}}.
\end{split}
\end{equation}
Here, we can define,
\begin{eqnarray}
\begin{aligned}
    \D_{\alpha\beta}^{r_k} = \bra{\Psi_{\alpha}}\frac{\partial}{\partial r_k}\ket{\Psi_{\beta}}, \\
    \D_{\alpha\beta}^{p_k} = \bra{\Psi_{\alpha}}\frac{\partial}{\partial p_k}\ket{\Psi_{\beta}}.
\end{aligned}
\end{eqnarray}
In this way, the hopping rate from surface $\alpha$ to surface $\beta$ in reciprocal space becomes,
\begin{equation}
    k_{\alpha\rightarrow \beta} = \Theta\left(-2\Re \left((p_k\cdot\D_{\alpha\beta}^{r_k}-\omega^2r_k\D_{\alpha\beta}^{p_k})\frac{\hat{\rho}^{{F(ad)}}_{\beta\alpha}}{\hat{\rho}^{{F(ad)}}_{\alpha\alpha}}\right)\right).
\end{equation}
After a hopping event, both position and momentum should be rescaled,
\begin{eqnarray}
\begin{aligned}
    p^{new} = p - \kappa \D_{\alpha\beta}^{r_k}/|\D_{\alpha\beta}^{r_k}|, \\
    r^{new} = r + \kappa \D_{\alpha\beta}^{p_k}/|\D_{\alpha\beta}^{p_k}|.
\end{aligned}
\end{eqnarray}

In Ref. \citenum{krotz2022reciprocal}, this surface hopping algorithm has been benchmarked against the numerically accurate hierarchical equations of motion (HEOM) method for calculating electronic dynamics in the solid state.

\section{Results and discussions}

In the following, we select a set of parameters to perform nonadiabatic dynamics in both real and reciprocal space. 
We set $J_1=0.3$, $J_2=0.6$, such that the band gap should be $0.6$.
Nuclear oscillation frequency $\omega=0.3$, temperature $kT=0.5$, reorganization energy $g^2\omega=0.075$.
We choose $10$ unit cell in Brillouin zone.
We set the lattice constant $a=1$.
For each trajectory, the initial condition of the phonon conforms to Boltzmann distribution as stated in Ref. \citenum{krotz2022reciprocal}. For the interband and intraband dynamics simulations, the electrons are concentrated at the $k=0$ point on the lower band at $t=0$. For the charge mobility simulation, they are concentrated at the central site in real space at $t=0$. 
To obtain these numerical results, we use a 4th-order Runge-Kutta algorithm to propagate both classical and quantum coordinates. The time step for the algorithm is set to $dt = 0.01$. We averaged $10,000$ trajectories to get the final dynamics results.

\begin{figure*}
    \centering
    \includegraphics[scale=0.1]{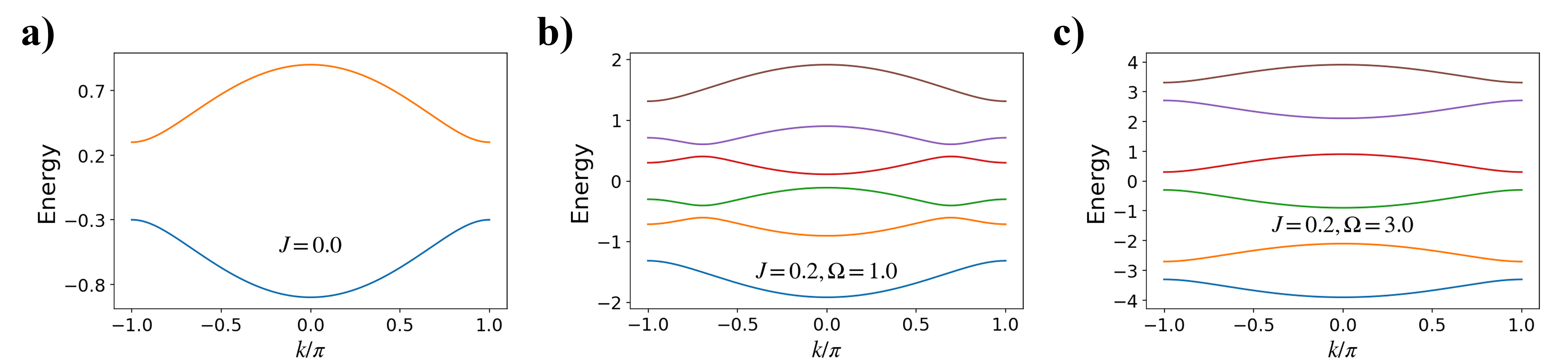}
    \caption{Electronic band structures under different drivings. a) Electronic band structure when there is no driving; b) Floquet electronic band structure when $J=0.2,\Omega=1.0$; c) Floquet electronic band structure when $J=0.2,\Omega=3.0$. }
    \label{fig:band_structure}
\end{figure*}

Figure \ref{fig:band_structure} shows the Floquet electronic bands under various driving conditions. In Floquet space, the original two bands are split into multiple Floquet replicas. At each step of the evolution, we transform the density matrix from Floquet space back to Hilbert space by summing the diagonal Floquet blocks, which represent the time-averaged driving effect.
Note that the electron-phonon coupling facilitates both interband and intraband transitions. After each time step evolution, we obtain the diabatic density matrix in reciprocal space. For each $k$-point, we then transform the diabatic density matrix to the adiabatic basis using the eigenvector without electron-phonon coupling to determine the electronic population in different $k$-points and bands.

\subsection{Interband and intraband transitions}

Interband transition is the electronic transitions between different energy bands in a material.
An interband transition occurs when an electron transitions from the valence band to the conduction band or vice versa. This process typically requires energy input or release equal to the energy gap (band gap) between the two bands.
Herein, the external periodic driving serves as a stimulus to facilitate electron transfer from the valence band to the conduction band.

\begin{figure*}
    \centering
    \includegraphics[scale=0.07]{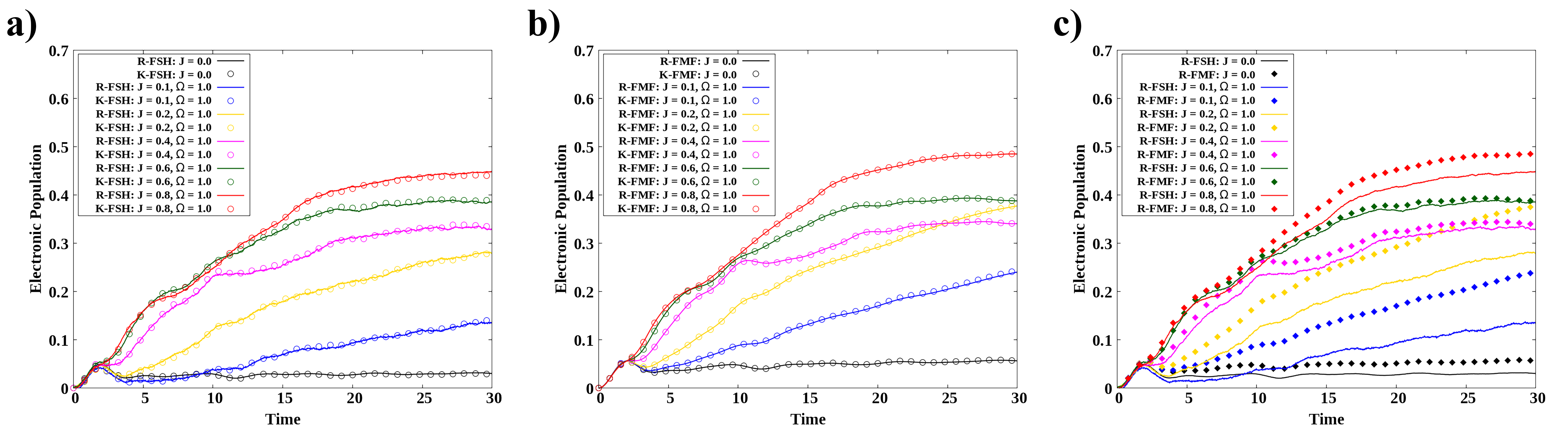}
    \caption{Interband transition (upper band electronic population) dynamics under different driving amplitudes $J$. a) Comparison of results between Floquet surface hopping algorithms in real space (R-FSH) and reciprocal space (K-FSH). Both methods show good agreement. Note that with increasing $J$, the interband transition increases. b) Comparison of results between Floquet mean field methods in real space (R-FMF) and reciprocal space (K-FMF). Both methods show good agreement. c) Comparison of results between FSH and FMF in real space. In the case of weak drivings, FMF shows a significant deviation from FSH.}
    \label{fig:diff_A_inter}
\end{figure*}

\begin{figure*}
    \centering
    \includegraphics[scale=0.5]{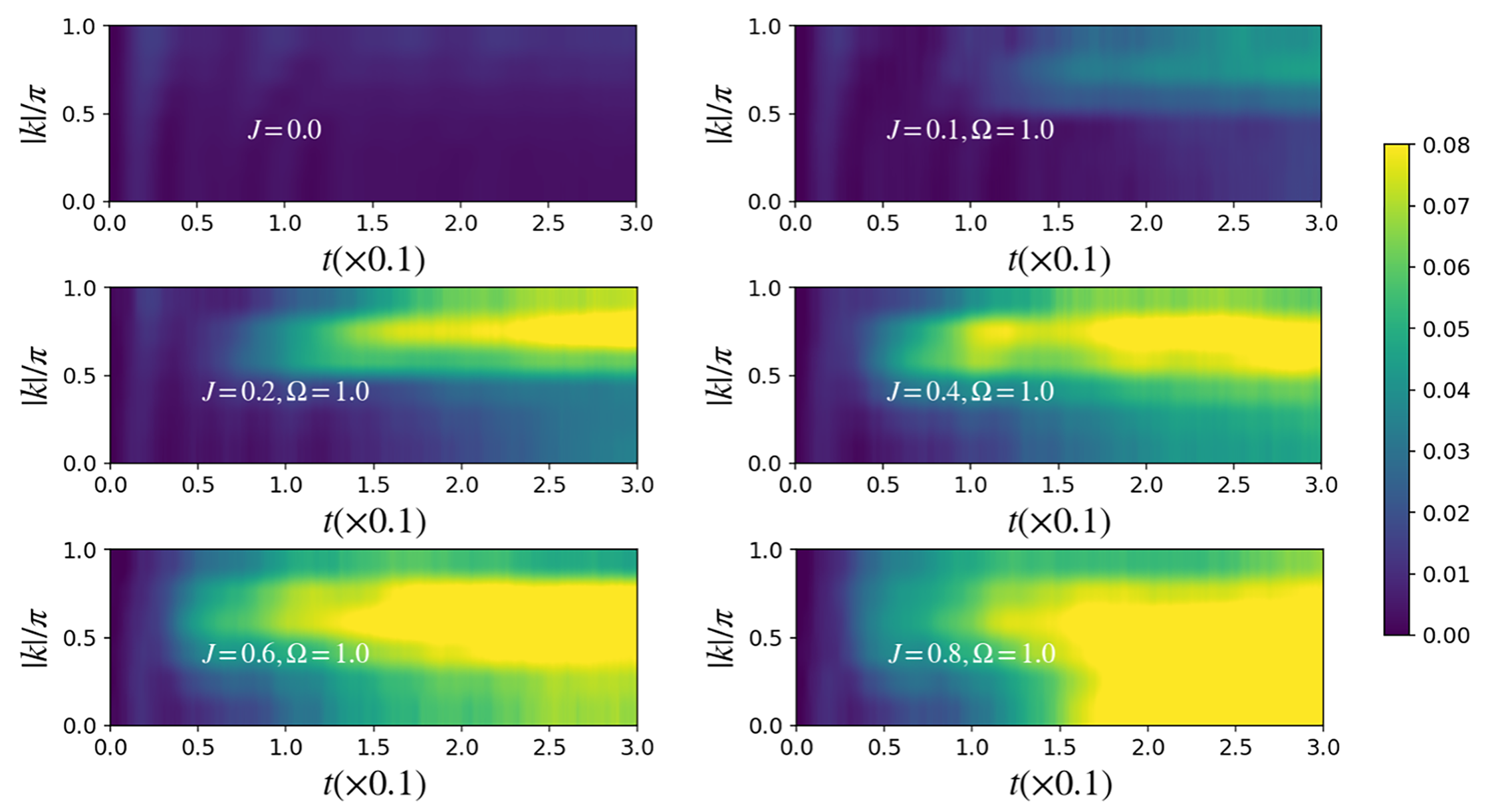}
    \caption{Heat map of the electronic population in the upper band under various driving amplitudes $J$, calculated by FSH in real space. Note that with increasing the driving amplitude $J$, the electron population at the upper band edge increases correspondingly.}
    \label{fig:heat_map_inter}
\end{figure*}

In Figure \ref{fig:diff_A_inter}, we show the interband dynamics under different driving amplitudes $J$ when fixing a driving frequency $\Omega=1.0$. Here, the electronic population refers to the total population on the upper band (i.e. conduction band). Figure \ref{fig:diff_A_inter}a shows the results from FSH in both real (solid line) and reciprocal spaces (hollow circle). We see a good agreement between these two algorithms.
When the amplitude of the driving is increased, we observe an enhancement of interband transitions from the valence band to the conduction band. This indicates that external drivings can facilitate interband transitions.
This phenomenon is also illustrated by the FMF method, as shown in Figure \ref{fig:diff_A_inter}b in both real and reciprocal space. However, significant deviations between FSH and FMF are observed when the driving amplitudes are small (weak drivings), as depicted in Figure \ref{fig:diff_A_inter}c.
The dynamics of the electronic population at different $k$ points in the upper band are shown in Figure \ref{fig:heat_map_inter}.
As the driving amplitude $J$ increases, more electrons are excited to the band edge of the upper band.

\begin{figure*}
    \centering
    \includegraphics[scale=0.07]{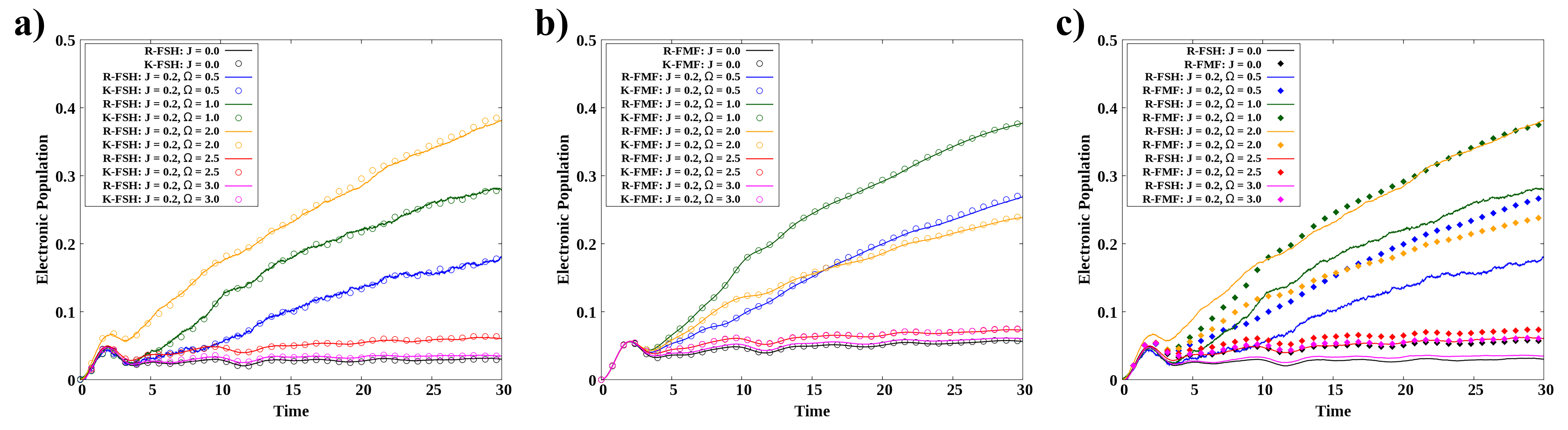}
    \caption{Interband transition (upper band electronic population) dynamics under different driving frequencies $\Omega$. a) Comparison of results between Floquet surface hopping algorithms in real space (R-FSH) and reciprocal space (K-FSH). Both methods show good agreement. Note that with increasing $\Omega$, the interband transition initially increases, then decreases. b) Comparison of results between Floquet mean field methods in real space (R-FMF) and reciprocal space (K-FMF). Both methods show good agreement. c) Comparison of results between FSH and FMF in real space. In the case of mediate drivings, FMF shows a significant deviation from FSH.}
    \label{fig:diff_W_inter}
\end{figure*}

\begin{figure*}
    \centering
    \includegraphics[scale=0.5]{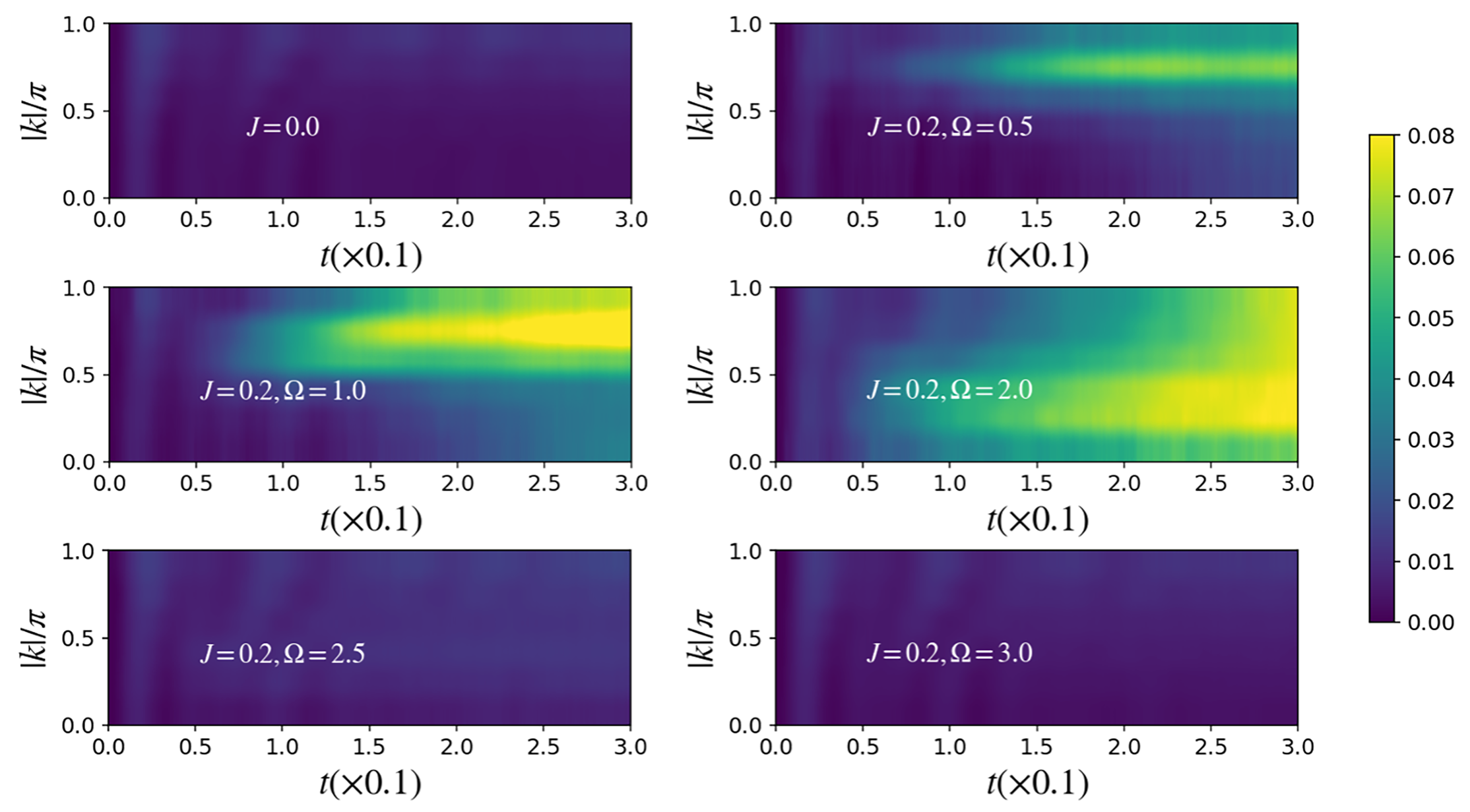}
    \caption{Heat map of the electronic population in the upper band under various driving frequencies $\Omega$, calculated by FSH in real space. Note that with increasing the driving frequency $\Omega$, the electron population at the upper band edge increases first, and then decreased.}
    \label{fig:heat_map_inter_W}
\end{figure*}

Next, we set a fixed driving amplitude $J=0.2$, and show the electronic dynamics on the upper band with the influence of different driving frequencies, as shown in Figure \ref{fig:diff_W_inter}. Figure \ref{fig:diff_W_inter}a presents results from FSH simulations, demonstrating that suitable driving frequencies facilitate interband transitions. As the driving frequency increases sufficiently (for example $\Omega=3.0$),  the system exhibits dynamics that resemble those observed in the absence of driving. This is because the system is unable to respond to fast drivings.
Same observations can also be present by FMF simulations in \ref{fig:diff_W_inter}b.
However, there are significant deviations between FSH and FMF under mediate driving frequencies.
The dynamics of the electronic population at different $k$ points in the upper band are shown in Figure \ref{fig:heat_map_inter_W}.
At appropriate driving frequencies $\Omega$, electron transitions from the lower band to the band edge of the upper band are increased.

\begin{figure*}
    \centering
    \includegraphics[scale=0.07]{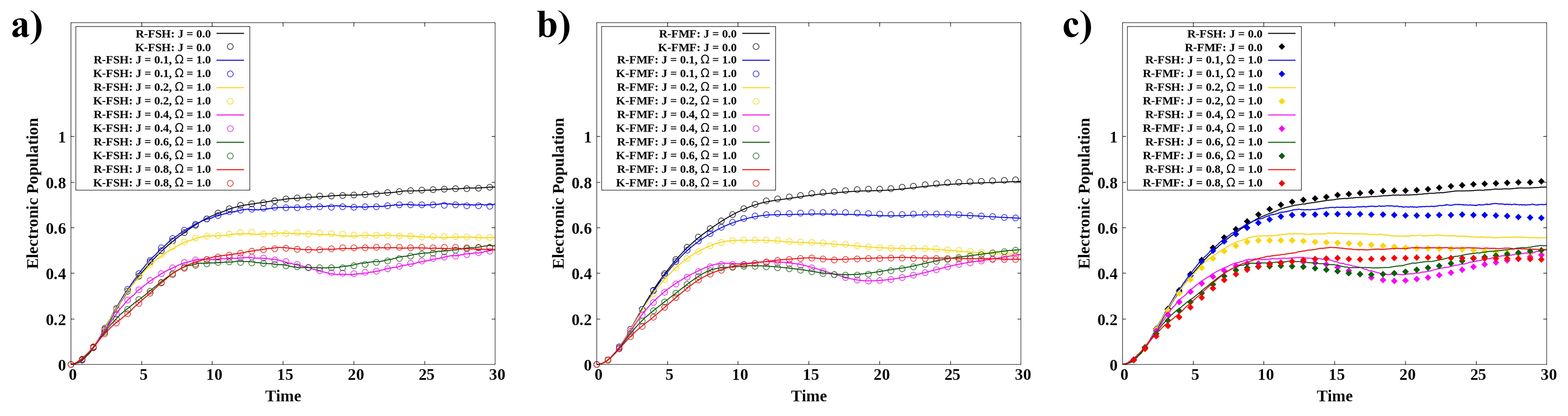}
    \caption{Intraband transition (lower band electronic population except $k=0$ point) dynamics under different driving amplitudes $J$. a) Comparison of results between Floquet surface hopping algorithms in real space (R-FSH) and reciprocal space (K-FSH). Both methods show good agreement. Note that with increasing $J$, the intraband transition decreases. b) Comparison of results between Floquet mean field methods in real space (R-FMF) and reciprocal space (K-FMF). Both methods show good agreement. c) Comparison of results between FSH and FMF in real space. In the case of weak drivings, FMF shows a significant deviation from FSH.}
    \label{fig:diff_A_intra}
\end{figure*}

\begin{figure*}
    \centering
    \includegraphics[scale=0.07]{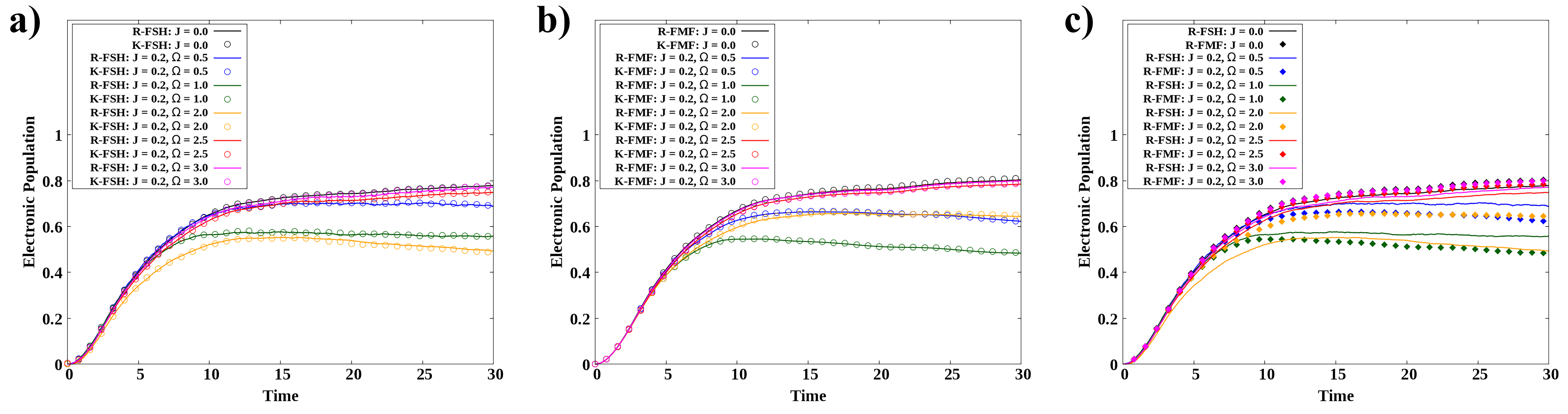}
    \caption{Intraband transition (lower band electronic population except $k=0$ point) dynamics under different driving frequencies $\Omega$. a) Comparison of results between Floquet surface hopping algorithms in real space (R-FSH) and reciprocal space (K-FSH). Both methods show good agreement. Note that with increasing $\Omega$, the intraband transition initially decreases, then increases. b) Comparison of results between Floquet mean field methods in real space (R-FMF) and reciprocal space (K-FMF). Both methods show good agreement. c) Comparison of results between FSH and FMF in real space. In the case of mediate drivings, FMF shows a significant deviation from FSH.}
    \label{fig:diff_W_intra}
\end{figure*}

Intraband transition refers to a quantum leap or transition of an electron within the same band of energy levels in a material.
In the presence of external drivings, we evaluate the electron population dynamics in the lower band excluding the point at $k=0$. We demonstrate that intraband transitions are suppressed, as illustrated in Figures \ref{fig:diff_A_intra} and \ref{fig:diff_W_intra}.
This observation is evident because we already know that periodic drivings increase interband transitions, thereby leaving fewer electrons available for intraband transitions.
Similarly, stronger drivings result in slower and less intraband transitions.
The system fails to respond to relatively fast drivings, resulting in dynamics similar to those observed in the absence of driving.

In summary, we introduce two semiclassical methods (FSH and FMF) to illustrate how external periodic drivings can promote interband transitions and correspondingly suppress intraband transitions.

\subsection{Charge mobility}

\begin{figure*}
    \centering
    \includegraphics[scale=0.08]{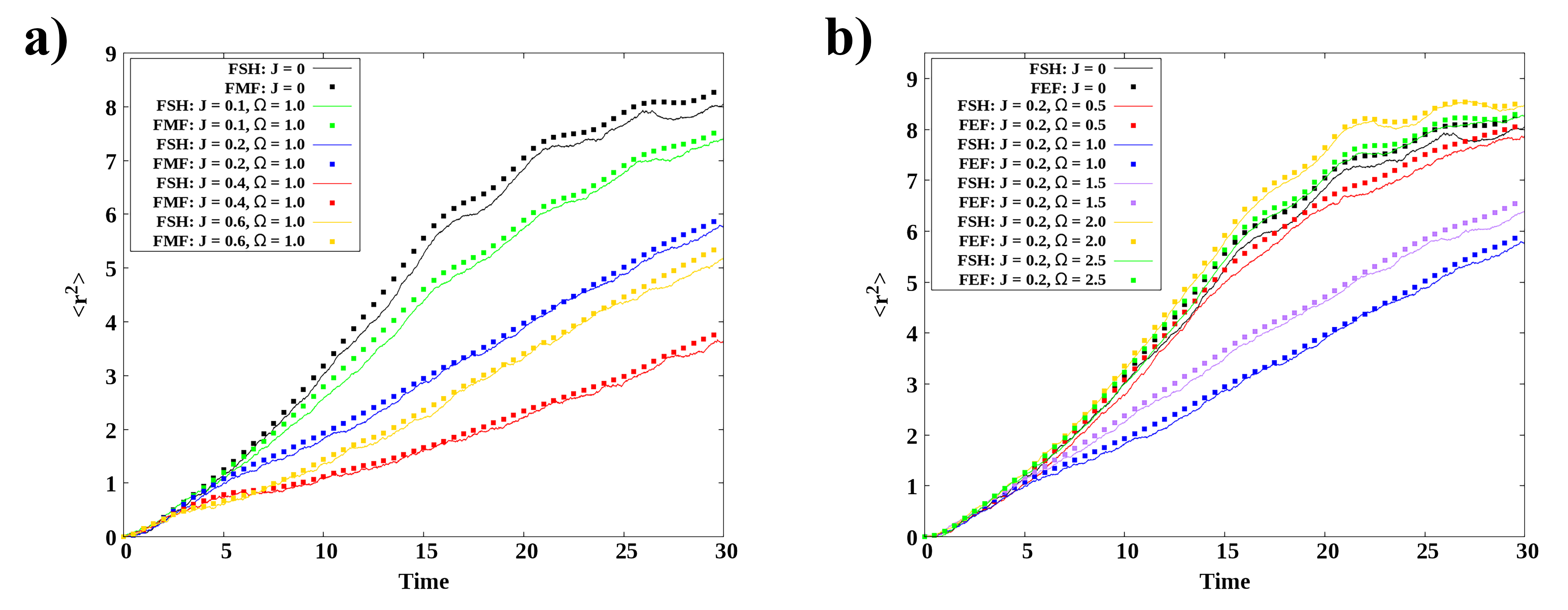}
    \caption{The dynamics of mean squared displacement $\langle r^2\rangle$ under the influence of different drivings. a) Comparison of results between FSH and FMF in real space when the driving frequency is fixed ($\Omega=1.0$). b) Comparison of results between FSH and FMF in real space when the driving amplitude is fixed ($J=0.2$). Note that periodic drivings exert a suppressing effect on charge mobility. }
    \label{fig:mobility}
\end{figure*}

Charge mobility is a crucial property in materials science and electronics because it determines how efficiently electrical current can flow.
Here, we calculate the mean squared displacement $\langle r^2\rangle$ under the influence of various drivings. In Figure \ref{fig:mobility} we can see that external periodic drivings exert a suppressing effect on charge mobility in this two-band model.
Remarkably, in real space, both the FSH and FMF methods produce identical dynamic results under different driving conditions. Due to its lower computational demand, employing the FMF method for charge mobility calculations provides a practical and efficient approach yielding reliable outcomes.

\section{Conclusions}

We have introduced two semiclassical algorithms, Floquet surface hopping (FSH) and Floquet mean field (FMF), to simulate carrier dynamics within a two-band model under periodic drivings. Both FSH and FMF methods are applicable in both real and reciprocal space. Our study demonstrates that periodic drivings efficiently enhance interband transitions while correspondingly suppressing intraband transitions. For simulations focusing on both interband and intraband transitions, the FSH method proves to be more reliable, having been previously benchmarked against the numerically accurate HEOM method in the one-band case. Furthermore, our results indicate that periodic drivings constrain charge mobility within this two-band model. For simulations of charge mobility under various periodic driving scenarios, FMF emerges as a practical and efficient approach.
Since external periodic drivings offer ways to manipulate and control the behavior of complex systems beyond their static equilibrium states, our proposed two methods here can play a significant role in understanding these dynamic effects. These approaches enhance our fundamental understanding of the systems under study and pave the way for optimizing their performance in diverse technological applications.

%\appendix

%\section{Appendixes}

\nocite{*}
\bibliography{aipsamp}% Produces the bibliography via BibTeX.

\begin{thebibliography}{43}%
\makeatletter
\providecommand \@ifxundefined [1]{%
 \@ifx{#1\undefined}
}%
\providecommand \@ifnum [1]{%
 \ifnum #1\expandafter \@firstoftwo
 \else \expandafter \@secondoftwo
 \fi
}%
\providecommand \@ifx [1]{%
 \ifx #1\expandafter \@firstoftwo
 \else \expandafter \@secondoftwo
 \fi
}%
\providecommand \natexlab [1]{#1}%
\providecommand \enquote  [1]{``#1''}%
\providecommand \bibnamefont  [1]{#1}%
\providecommand \bibfnamefont [1]{#1}%
\providecommand \citenamefont [1]{#1}%
\providecommand \href@noop [0]{\@secondoftwo}%
\providecommand \href [0]{\begingroup \@sanitize@url \@href}%
\providecommand \@href[1]{\@@startlink{#1}\@@href}%
\providecommand \@@href[1]{\endgroup#1\@@endlink}%
\providecommand \@sanitize@url [0]{\catcode `\\12\catcode `\$12\catcode
  `\&12\catcode `\#12\catcode `\^12\catcode `\_12\catcode `\%12\relax}%
\providecommand \@@startlink[1]{}%
\providecommand \@@endlink[0]{}%
\providecommand \url  [0]{\begingroup\@sanitize@url \@url }%
\providecommand \@url [1]{\endgroup\@href {#1}{\urlprefix }}%
\providecommand \urlprefix  [0]{URL }%
\providecommand \Eprint [0]{\href }%
\providecommand \doibase [0]{http://dx.doi.org/}%
\providecommand \selectlanguage [0]{\@gobble}%
\providecommand \bibinfo  [0]{\@secondoftwo}%
\providecommand \bibfield  [0]{\@secondoftwo}%
\providecommand \translation [1]{[#1]}%
\providecommand \BibitemOpen [0]{}%
\providecommand \bibitemStop [0]{}%
\providecommand \bibitemNoStop [0]{.\EOS\space}%
\providecommand \EOS [0]{\spacefactor3000\relax}%
\providecommand \BibitemShut  [1]{\csname bibitem#1\endcsname}%
\let\auto@bib@innerbib\@empty
%</preamble>
\bibitem [{\citenamefont {Ulbricht}\ \emph {et~al.}(2011)\citenamefont
  {Ulbricht}, \citenamefont {Hendry}, \citenamefont {Shan}, \citenamefont
  {Heinz},\ and\ \citenamefont {Bonn}}]{ulbricht2011carrier}%
  \BibitemOpen
  \bibfield  {author} {\bibinfo {author} {\bibfnamefont {R.}~\bibnamefont
  {Ulbricht}}, \bibinfo {author} {\bibfnamefont {E.}~\bibnamefont {Hendry}},
  \bibinfo {author} {\bibfnamefont {J.}~\bibnamefont {Shan}}, \bibinfo {author}
  {\bibfnamefont {T.~F.}\ \bibnamefont {Heinz}}, \ and\ \bibinfo {author}
  {\bibfnamefont {M.}~\bibnamefont {Bonn}},\ }\bibfield  {title} {\enquote
  {\bibinfo {title} {Carrier dynamics in semiconductors studied with
  time-resolved terahertz spectroscopy},}\ }\href@noop {} {\bibfield  {journal}
  {\bibinfo  {journal} {Reviews of Modern Physics}\ }\textbf {\bibinfo {volume}
  {83}},\ \bibinfo {pages} {543} (\bibinfo {year} {2011})}\BibitemShut
  {NoStop}%
\bibitem [{\citenamefont {Liao}\ \emph {et~al.}(2020)\citenamefont {Liao},
  \citenamefont {Wu}, \citenamefont {Jiang}, \citenamefont {Zhong},
  \citenamefont {Wang},\ and\ \citenamefont {Kuang}}]{liao2020understanding}%
  \BibitemOpen
  \bibfield  {author} {\bibinfo {author} {\bibfnamefont {J.-F.}\ \bibnamefont
  {Liao}}, \bibinfo {author} {\bibfnamefont {W.-Q.}\ \bibnamefont {Wu}},
  \bibinfo {author} {\bibfnamefont {Y.}~\bibnamefont {Jiang}}, \bibinfo
  {author} {\bibfnamefont {J.-X.}\ \bibnamefont {Zhong}}, \bibinfo {author}
  {\bibfnamefont {L.}~\bibnamefont {Wang}}, \ and\ \bibinfo {author}
  {\bibfnamefont {D.-B.}\ \bibnamefont {Kuang}},\ }\bibfield  {title} {\enquote
  {\bibinfo {title} {Understanding of carrier dynamics, heterojunction merits
  and device physics: towards designing efficient carrier transport layer-free
  perovskite solar cells},}\ }\href@noop {} {\bibfield  {journal} {\bibinfo
  {journal} {Chemical Society Reviews}\ }\textbf {\bibinfo {volume} {49}},\
  \bibinfo {pages} {354--381} (\bibinfo {year} {2020})}\BibitemShut {NoStop}%
\bibitem [{\citenamefont {Zhu}\ \emph {et~al.}(2019)\citenamefont {Zhu},
  \citenamefont {Niu}, \citenamefont {Fu}, \citenamefont {Li}, \citenamefont
  {Hu}, \citenamefont {Chen}, \citenamefont {He}, \citenamefont {Na},
  \citenamefont {Liu}, \citenamefont {Zai} \emph {et~al.}}]{zhu2019strain}%
  \BibitemOpen
  \bibfield  {author} {\bibinfo {author} {\bibfnamefont {C.}~\bibnamefont
  {Zhu}}, \bibinfo {author} {\bibfnamefont {X.}~\bibnamefont {Niu}}, \bibinfo
  {author} {\bibfnamefont {Y.}~\bibnamefont {Fu}}, \bibinfo {author}
  {\bibfnamefont {N.}~\bibnamefont {Li}}, \bibinfo {author} {\bibfnamefont
  {C.}~\bibnamefont {Hu}}, \bibinfo {author} {\bibfnamefont {Y.}~\bibnamefont
  {Chen}}, \bibinfo {author} {\bibfnamefont {X.}~\bibnamefont {He}}, \bibinfo
  {author} {\bibfnamefont {G.}~\bibnamefont {Na}}, \bibinfo {author}
  {\bibfnamefont {P.}~\bibnamefont {Liu}}, \bibinfo {author} {\bibfnamefont
  {H.}~\bibnamefont {Zai}},  \emph {et~al.},\ }\bibfield  {title} {\enquote
  {\bibinfo {title} {Strain engineering in perovskite solar cells and its
  impacts on carrier dynamics},}\ }\href@noop {} {\bibfield  {journal}
  {\bibinfo  {journal} {Nature communications}\ }\textbf {\bibinfo {volume}
  {10}},\ \bibinfo {pages} {815} (\bibinfo {year} {2019})}\BibitemShut
  {NoStop}%
\bibitem [{\citenamefont {Quan}, \citenamefont {Yue},\ and\ \citenamefont
  {Liao}(2021)}]{quan2021impact}%
  \BibitemOpen
  \bibfield  {author} {\bibinfo {author} {\bibfnamefont {Y.}~\bibnamefont
  {Quan}}, \bibinfo {author} {\bibfnamefont {S.}~\bibnamefont {Yue}}, \ and\
  \bibinfo {author} {\bibfnamefont {B.}~\bibnamefont {Liao}},\ }\bibfield
  {title} {\enquote {\bibinfo {title} {Impact of electron-phonon interaction on
  thermal transport: A review},}\ }\href@noop {} {\bibfield  {journal}
  {\bibinfo  {journal} {Nanoscale and Microscale Thermophysical Engineering}\
  }\textbf {\bibinfo {volume} {25}},\ \bibinfo {pages} {73--90} (\bibinfo
  {year} {2021})}\BibitemShut {NoStop}%
\bibitem [{\citenamefont {Giustino}(2017)}]{giustino2017electron}%
  \BibitemOpen
  \bibfield  {author} {\bibinfo {author} {\bibfnamefont {F.}~\bibnamefont
  {Giustino}},\ }\bibfield  {title} {\enquote {\bibinfo {title}
  {Electron-phonon interactions from first principles},}\ }\href@noop {}
  {\bibfield  {journal} {\bibinfo  {journal} {Reviews of Modern Physics}\
  }\textbf {\bibinfo {volume} {89}},\ \bibinfo {pages} {015003} (\bibinfo
  {year} {2017})}\BibitemShut {NoStop}%
\bibitem [{\citenamefont {Ghosh}\ \emph {et~al.}(2017)\citenamefont {Ghosh},
  \citenamefont {Aharon}, \citenamefont {Etgar},\ and\ \citenamefont
  {Ruhman}}]{ghosh2017free}%
  \BibitemOpen
  \bibfield  {author} {\bibinfo {author} {\bibfnamefont {T.}~\bibnamefont
  {Ghosh}}, \bibinfo {author} {\bibfnamefont {S.}~\bibnamefont {Aharon}},
  \bibinfo {author} {\bibfnamefont {L.}~\bibnamefont {Etgar}}, \ and\ \bibinfo
  {author} {\bibfnamefont {S.}~\bibnamefont {Ruhman}},\ }\bibfield  {title}
  {\enquote {\bibinfo {title} {Free carrier emergence and onset of
  electron--phonon coupling in methylammonium lead halide perovskite films},}\
  }\href@noop {} {\bibfield  {journal} {\bibinfo  {journal} {Journal of the
  American Chemical Society}\ }\textbf {\bibinfo {volume} {139}},\ \bibinfo
  {pages} {18262--18270} (\bibinfo {year} {2017})}\BibitemShut {NoStop}%
\bibitem [{\citenamefont {Wright}\ \emph {et~al.}(2016)\citenamefont {Wright},
  \citenamefont {Verdi}, \citenamefont {Milot}, \citenamefont {Eperon},
  \citenamefont {P{\'e}rez-Osorio}, \citenamefont {Snaith}, \citenamefont
  {Giustino}, \citenamefont {Johnston},\ and\ \citenamefont
  {Herz}}]{wright2016electron}%
  \BibitemOpen
  \bibfield  {author} {\bibinfo {author} {\bibfnamefont {A.~D.}\ \bibnamefont
  {Wright}}, \bibinfo {author} {\bibfnamefont {C.}~\bibnamefont {Verdi}},
  \bibinfo {author} {\bibfnamefont {R.~L.}\ \bibnamefont {Milot}}, \bibinfo
  {author} {\bibfnamefont {G.~E.}\ \bibnamefont {Eperon}}, \bibinfo {author}
  {\bibfnamefont {M.~A.}\ \bibnamefont {P{\'e}rez-Osorio}}, \bibinfo {author}
  {\bibfnamefont {H.~J.}\ \bibnamefont {Snaith}}, \bibinfo {author}
  {\bibfnamefont {F.}~\bibnamefont {Giustino}}, \bibinfo {author}
  {\bibfnamefont {M.~B.}\ \bibnamefont {Johnston}}, \ and\ \bibinfo {author}
  {\bibfnamefont {L.~M.}\ \bibnamefont {Herz}},\ }\bibfield  {title} {\enquote
  {\bibinfo {title} {Electron--phonon coupling in hybrid lead halide
  perovskites},}\ }\href@noop {} {\bibfield  {journal} {\bibinfo  {journal}
  {Nature communications}\ }\textbf {\bibinfo {volume} {7}},\ \bibinfo {pages}
  {11755} (\bibinfo {year} {2016})}\BibitemShut {NoStop}%
\bibitem [{\citenamefont {Wehrenberg}, \citenamefont {Wang},\ and\
  \citenamefont {Guyot-Sionnest}(2002)}]{wehrenberg2002interband}%
  \BibitemOpen
  \bibfield  {author} {\bibinfo {author} {\bibfnamefont {B.~L.}\ \bibnamefont
  {Wehrenberg}}, \bibinfo {author} {\bibfnamefont {C.}~\bibnamefont {Wang}}, \
  and\ \bibinfo {author} {\bibfnamefont {P.}~\bibnamefont {Guyot-Sionnest}},\
  }\bibfield  {title} {\enquote {\bibinfo {title} {Interband and intraband
  optical studies of pbse colloidal quantum dots},}\ }\href@noop {} {\bibfield
  {journal} {\bibinfo  {journal} {The Journal of Physical Chemistry B}\
  }\textbf {\bibinfo {volume} {106}},\ \bibinfo {pages} {10634--10640}
  (\bibinfo {year} {2002})}\BibitemShut {NoStop}%
\bibitem [{\citenamefont {Wang}\ \emph {et~al.}(2010)\citenamefont {Wang},
  \citenamefont {Nan}, \citenamefont {Yang}, \citenamefont {Peng},
  \citenamefont {Li},\ and\ \citenamefont {Shuai}}]{wang2010computational}%
  \BibitemOpen
  \bibfield  {author} {\bibinfo {author} {\bibfnamefont {L.}~\bibnamefont
  {Wang}}, \bibinfo {author} {\bibfnamefont {G.}~\bibnamefont {Nan}}, \bibinfo
  {author} {\bibfnamefont {X.}~\bibnamefont {Yang}}, \bibinfo {author}
  {\bibfnamefont {Q.}~\bibnamefont {Peng}}, \bibinfo {author} {\bibfnamefont
  {Q.}~\bibnamefont {Li}}, \ and\ \bibinfo {author} {\bibfnamefont
  {Z.}~\bibnamefont {Shuai}},\ }\bibfield  {title} {\enquote {\bibinfo {title}
  {Computational methods for design of organic materials with high charge
  mobility},}\ }\href@noop {} {\bibfield  {journal} {\bibinfo  {journal}
  {Chemical Society Reviews}\ }\textbf {\bibinfo {volume} {39}},\ \bibinfo
  {pages} {423--434} (\bibinfo {year} {2010})}\BibitemShut {NoStop}%
\bibitem [{\citenamefont {Creffield}(2007)}]{creffield2007quantum}%
  \BibitemOpen
  \bibfield  {author} {\bibinfo {author} {\bibfnamefont {C.}~\bibnamefont
  {Creffield}},\ }\bibfield  {title} {\enquote {\bibinfo {title} {Quantum
  control and entanglement using periodic driving fields},}\ }\href@noop {}
  {\bibfield  {journal} {\bibinfo  {journal} {Physical review letters}\
  }\textbf {\bibinfo {volume} {99}},\ \bibinfo {pages} {110501} (\bibinfo
  {year} {2007})}\BibitemShut {NoStop}%
\bibitem [{\citenamefont {Bordia}\ \emph {et~al.}(2017)\citenamefont {Bordia},
  \citenamefont {L{\"u}schen}, \citenamefont {Schneider}, \citenamefont
  {Knap},\ and\ \citenamefont {Bloch}}]{bordia2017periodically}%
  \BibitemOpen
  \bibfield  {author} {\bibinfo {author} {\bibfnamefont {P.}~\bibnamefont
  {Bordia}}, \bibinfo {author} {\bibfnamefont {H.}~\bibnamefont {L{\"u}schen}},
  \bibinfo {author} {\bibfnamefont {U.}~\bibnamefont {Schneider}}, \bibinfo
  {author} {\bibfnamefont {M.}~\bibnamefont {Knap}}, \ and\ \bibinfo {author}
  {\bibfnamefont {I.}~\bibnamefont {Bloch}},\ }\bibfield  {title} {\enquote
  {\bibinfo {title} {Periodically driving a many-body localized quantum
  system},}\ }\href@noop {} {\bibfield  {journal} {\bibinfo  {journal} {Nature
  Physics}\ }\textbf {\bibinfo {volume} {13}},\ \bibinfo {pages} {460--464}
  (\bibinfo {year} {2017})}\BibitemShut {NoStop}%
\bibitem [{\citenamefont {Kim}\ \emph {et~al.}(2018)\citenamefont {Kim},
  \citenamefont {Bewley}, \citenamefont {Merritt}, \citenamefont {Canedy},
  \citenamefont {Warren}, \citenamefont {Vurgaftman}, \citenamefont {Meyer},\
  and\ \citenamefont {Kim}}]{kim2018improved}%
  \BibitemOpen
  \bibfield  {author} {\bibinfo {author} {\bibfnamefont {C.~S.}\ \bibnamefont
  {Kim}}, \bibinfo {author} {\bibfnamefont {W.~W.}\ \bibnamefont {Bewley}},
  \bibinfo {author} {\bibfnamefont {C.~D.}\ \bibnamefont {Merritt}}, \bibinfo
  {author} {\bibfnamefont {C.~L.}\ \bibnamefont {Canedy}}, \bibinfo {author}
  {\bibfnamefont {M.~V.}\ \bibnamefont {Warren}}, \bibinfo {author}
  {\bibfnamefont {I.}~\bibnamefont {Vurgaftman}}, \bibinfo {author}
  {\bibfnamefont {J.~R.}\ \bibnamefont {Meyer}}, \ and\ \bibinfo {author}
  {\bibfnamefont {M.}~\bibnamefont {Kim}},\ }\bibfield  {title} {\enquote
  {\bibinfo {title} {Improved mid-infrared interband cascade light-emitting
  devices},}\ }\href@noop {} {\bibfield  {journal} {\bibinfo  {journal}
  {Optical Engineering}\ }\textbf {\bibinfo {volume} {57}},\ \bibinfo {pages}
  {011002--011002} (\bibinfo {year} {2018})}\BibitemShut {NoStop}%
\bibitem [{\citenamefont {Meyer}\ \emph {et~al.}(2020)\citenamefont {Meyer},
  \citenamefont {Bewley}, \citenamefont {Canedy}, \citenamefont {Kim},
  \citenamefont {Kim}, \citenamefont {Merritt},\ and\ \citenamefont
  {Vurgaftman}}]{meyer2020interband}%
  \BibitemOpen
  \bibfield  {author} {\bibinfo {author} {\bibfnamefont {J.~R.}\ \bibnamefont
  {Meyer}}, \bibinfo {author} {\bibfnamefont {W.~W.}\ \bibnamefont {Bewley}},
  \bibinfo {author} {\bibfnamefont {C.~L.}\ \bibnamefont {Canedy}}, \bibinfo
  {author} {\bibfnamefont {C.~S.}\ \bibnamefont {Kim}}, \bibinfo {author}
  {\bibfnamefont {M.}~\bibnamefont {Kim}}, \bibinfo {author} {\bibfnamefont
  {C.~D.}\ \bibnamefont {Merritt}}, \ and\ \bibinfo {author} {\bibfnamefont
  {I.}~\bibnamefont {Vurgaftman}},\ }\bibfield  {title} {\enquote {\bibinfo
  {title} {The interband cascade laser},}\ }in\ \href@noop {} {\emph {\bibinfo
  {booktitle} {Photonics}}},\ Vol.~\bibinfo {volume} {7}\ (\bibinfo
  {organization} {MDPI},\ \bibinfo {year} {2020})\ p.~\bibinfo {pages}
  {75}\BibitemShut {NoStop}%
\bibitem [{\citenamefont {Wu}\ \emph {et~al.}(2012)\citenamefont {Wu},
  \citenamefont {Makableh}, \citenamefont {Vasan}, \citenamefont {Manasreh},
  \citenamefont {Liang}, \citenamefont {Reyner},\ and\ \citenamefont
  {Huffaker}}]{wu2012strong}%
  \BibitemOpen
  \bibfield  {author} {\bibinfo {author} {\bibfnamefont {J.}~\bibnamefont
  {Wu}}, \bibinfo {author} {\bibfnamefont {Y.}~\bibnamefont {Makableh}},
  \bibinfo {author} {\bibfnamefont {R.}~\bibnamefont {Vasan}}, \bibinfo
  {author} {\bibfnamefont {M.}~\bibnamefont {Manasreh}}, \bibinfo {author}
  {\bibfnamefont {B.}~\bibnamefont {Liang}}, \bibinfo {author} {\bibfnamefont
  {C.}~\bibnamefont {Reyner}}, \ and\ \bibinfo {author} {\bibfnamefont
  {D.}~\bibnamefont {Huffaker}},\ }\bibfield  {title} {\enquote {\bibinfo
  {title} {Strong interband transitions in inas quantum dots solar cell},}\
  }\href@noop {} {\bibfield  {journal} {\bibinfo  {journal} {Applied Physics
  Letters}\ }\textbf {\bibinfo {volume} {100}} (\bibinfo {year}
  {2012})}\BibitemShut {NoStop}%
\bibitem [{\citenamefont {Okada}\ \emph {et~al.}(2015)\citenamefont {Okada},
  \citenamefont {Ekins-Daukes}, \citenamefont {Kita}, \citenamefont {Tamaki},
  \citenamefont {Yoshida}, \citenamefont {Pusch}, \citenamefont {Hess},
  \citenamefont {Phillips}, \citenamefont {Farrell}, \citenamefont {Yoshida}
  \emph {et~al.}}]{okada2015intermediate}%
  \BibitemOpen
  \bibfield  {author} {\bibinfo {author} {\bibfnamefont {Y.}~\bibnamefont
  {Okada}}, \bibinfo {author} {\bibfnamefont {N.}~\bibnamefont {Ekins-Daukes}},
  \bibinfo {author} {\bibfnamefont {T.}~\bibnamefont {Kita}}, \bibinfo {author}
  {\bibfnamefont {R.}~\bibnamefont {Tamaki}}, \bibinfo {author} {\bibfnamefont
  {M.}~\bibnamefont {Yoshida}}, \bibinfo {author} {\bibfnamefont
  {A.}~\bibnamefont {Pusch}}, \bibinfo {author} {\bibfnamefont
  {O.}~\bibnamefont {Hess}}, \bibinfo {author} {\bibfnamefont {C.}~\bibnamefont
  {Phillips}}, \bibinfo {author} {\bibfnamefont {D.}~\bibnamefont {Farrell}},
  \bibinfo {author} {\bibfnamefont {K.}~\bibnamefont {Yoshida}},  \emph
  {et~al.},\ }\bibfield  {title} {\enquote {\bibinfo {title} {Intermediate band
  solar cells: Recent progress and future directions},}\ }\href@noop {}
  {\bibfield  {journal} {\bibinfo  {journal} {Applied physics reviews}\
  }\textbf {\bibinfo {volume} {2}} (\bibinfo {year} {2015})}\BibitemShut
  {NoStop}%
\bibitem [{\citenamefont {Grancini}\ \emph {et~al.}(2013)\citenamefont
  {Grancini}, \citenamefont {Maiuri}, \citenamefont {Fazzi}, \citenamefont
  {Petrozza}, \citenamefont {Egelhaaf}, \citenamefont {Brida}, \citenamefont
  {Cerullo},\ and\ \citenamefont {Lanzani}}]{grancini2013hot}%
  \BibitemOpen
  \bibfield  {author} {\bibinfo {author} {\bibfnamefont {G.}~\bibnamefont
  {Grancini}}, \bibinfo {author} {\bibfnamefont {M.}~\bibnamefont {Maiuri}},
  \bibinfo {author} {\bibfnamefont {D.}~\bibnamefont {Fazzi}}, \bibinfo
  {author} {\bibfnamefont {A.}~\bibnamefont {Petrozza}}, \bibinfo {author}
  {\bibfnamefont {H.-J.}\ \bibnamefont {Egelhaaf}}, \bibinfo {author}
  {\bibfnamefont {D.}~\bibnamefont {Brida}}, \bibinfo {author} {\bibfnamefont
  {G.}~\bibnamefont {Cerullo}}, \ and\ \bibinfo {author} {\bibfnamefont
  {G.}~\bibnamefont {Lanzani}},\ }\bibfield  {title} {\enquote {\bibinfo
  {title} {Hot exciton dissociation in polymer solar cells},}\ }\href@noop {}
  {\bibfield  {journal} {\bibinfo  {journal} {Nature materials}\ }\textbf
  {\bibinfo {volume} {12}},\ \bibinfo {pages} {29--33} (\bibinfo {year}
  {2013})}\BibitemShut {NoStop}%
\bibitem [{\citenamefont {G{\'o}mez-Le{\'o}n}\ and\ \citenamefont
  {Platero}(2013)}]{gomez2013floquet}%
  \BibitemOpen
  \bibfield  {author} {\bibinfo {author} {\bibfnamefont {A.}~\bibnamefont
  {G{\'o}mez-Le{\'o}n}}\ and\ \bibinfo {author} {\bibfnamefont
  {G.}~\bibnamefont {Platero}},\ }\bibfield  {title} {\enquote {\bibinfo
  {title} {Floquet-bloch theory and topology in periodically driven
  lattices},}\ }\href@noop {} {\bibfield  {journal} {\bibinfo  {journal}
  {Physical review letters}\ }\textbf {\bibinfo {volume} {110}},\ \bibinfo
  {pages} {200403} (\bibinfo {year} {2013})}\BibitemShut {NoStop}%
\bibitem [{\citenamefont {Wang}\ \emph {et~al.}(2013)\citenamefont {Wang},
  \citenamefont {Steinberg}, \citenamefont {Jarillo-Herrero},\ and\
  \citenamefont {Gedik}}]{wang2013observation}%
  \BibitemOpen
  \bibfield  {author} {\bibinfo {author} {\bibfnamefont {Y.}~\bibnamefont
  {Wang}}, \bibinfo {author} {\bibfnamefont {H.}~\bibnamefont {Steinberg}},
  \bibinfo {author} {\bibfnamefont {P.}~\bibnamefont {Jarillo-Herrero}}, \ and\
  \bibinfo {author} {\bibfnamefont {N.}~\bibnamefont {Gedik}},\ }\bibfield
  {title} {\enquote {\bibinfo {title} {Observation of floquet-bloch states on
  the surface of a topological insulator},}\ }\href@noop {} {\bibfield
  {journal} {\bibinfo  {journal} {Science}\ }\textbf {\bibinfo {volume}
  {342}},\ \bibinfo {pages} {453--457} (\bibinfo {year} {2013})}\BibitemShut
  {NoStop}%
\bibitem [{\citenamefont {Upreti}\ \emph {et~al.}(2020)\citenamefont {Upreti},
  \citenamefont {Evain}, \citenamefont {Randoux}, \citenamefont {Suret},
  \citenamefont {Amo},\ and\ \citenamefont {Delplace}}]{upreti2020topological}%
  \BibitemOpen
  \bibfield  {author} {\bibinfo {author} {\bibfnamefont {L.~K.}\ \bibnamefont
  {Upreti}}, \bibinfo {author} {\bibfnamefont {C.}~\bibnamefont {Evain}},
  \bibinfo {author} {\bibfnamefont {S.}~\bibnamefont {Randoux}}, \bibinfo
  {author} {\bibfnamefont {P.}~\bibnamefont {Suret}}, \bibinfo {author}
  {\bibfnamefont {A.}~\bibnamefont {Amo}}, \ and\ \bibinfo {author}
  {\bibfnamefont {P.}~\bibnamefont {Delplace}},\ }\bibfield  {title} {\enquote
  {\bibinfo {title} {Topological swing of bloch oscillations in quantum
  walks},}\ }\href@noop {} {\bibfield  {journal} {\bibinfo  {journal} {Physical
  Review Letters}\ }\textbf {\bibinfo {volume} {125}},\ \bibinfo {pages}
  {186804} (\bibinfo {year} {2020})}\BibitemShut {NoStop}%
\bibitem [{\citenamefont {Lucchini}\ \emph {et~al.}(2022)\citenamefont
  {Lucchini}, \citenamefont {Medeghini}, \citenamefont {Wu}, \citenamefont
  {Vismarra}, \citenamefont {Borrego-Varillas}, \citenamefont {Crego},
  \citenamefont {Frassetto}, \citenamefont {Poletto}, \citenamefont {Sato},
  \citenamefont {H{\"u}bener} \emph {et~al.}}]{lucchini2022controlling}%
  \BibitemOpen
  \bibfield  {author} {\bibinfo {author} {\bibfnamefont {M.}~\bibnamefont
  {Lucchini}}, \bibinfo {author} {\bibfnamefont {F.}~\bibnamefont {Medeghini}},
  \bibinfo {author} {\bibfnamefont {Y.}~\bibnamefont {Wu}}, \bibinfo {author}
  {\bibfnamefont {F.}~\bibnamefont {Vismarra}}, \bibinfo {author}
  {\bibfnamefont {R.}~\bibnamefont {Borrego-Varillas}}, \bibinfo {author}
  {\bibfnamefont {A.}~\bibnamefont {Crego}}, \bibinfo {author} {\bibfnamefont
  {F.}~\bibnamefont {Frassetto}}, \bibinfo {author} {\bibfnamefont
  {L.}~\bibnamefont {Poletto}}, \bibinfo {author} {\bibfnamefont {S.~A.}\
  \bibnamefont {Sato}}, \bibinfo {author} {\bibfnamefont {H.}~\bibnamefont
  {H{\"u}bener}},  \emph {et~al.},\ }\bibfield  {title} {\enquote {\bibinfo
  {title} {Controlling floquet states on ultrashort time scales},}\ }\href@noop
  {} {\bibfield  {journal} {\bibinfo  {journal} {Nature Communications}\
  }\textbf {\bibinfo {volume} {13}},\ \bibinfo {pages} {7103} (\bibinfo {year}
  {2022})}\BibitemShut {NoStop}%
\bibitem [{\citenamefont {Nakagawa}\ \emph {et~al.}(2022)\citenamefont
  {Nakagawa}, \citenamefont {Hirori}, \citenamefont {Sato}, \citenamefont
  {Tahara}, \citenamefont {Sekiguchi}, \citenamefont {Yumoto}, \citenamefont
  {Saruyama}, \citenamefont {Sato}, \citenamefont {Teranishi},\ and\
  \citenamefont {Kanemitsu}}]{nakagawa2022size}%
  \BibitemOpen
  \bibfield  {author} {\bibinfo {author} {\bibfnamefont {K.}~\bibnamefont
  {Nakagawa}}, \bibinfo {author} {\bibfnamefont {H.}~\bibnamefont {Hirori}},
  \bibinfo {author} {\bibfnamefont {S.~A.}\ \bibnamefont {Sato}}, \bibinfo
  {author} {\bibfnamefont {H.}~\bibnamefont {Tahara}}, \bibinfo {author}
  {\bibfnamefont {F.}~\bibnamefont {Sekiguchi}}, \bibinfo {author}
  {\bibfnamefont {G.}~\bibnamefont {Yumoto}}, \bibinfo {author} {\bibfnamefont
  {M.}~\bibnamefont {Saruyama}}, \bibinfo {author} {\bibfnamefont
  {R.}~\bibnamefont {Sato}}, \bibinfo {author} {\bibfnamefont {T.}~\bibnamefont
  {Teranishi}}, \ and\ \bibinfo {author} {\bibfnamefont {Y.}~\bibnamefont
  {Kanemitsu}},\ }\bibfield  {title} {\enquote {\bibinfo {title}
  {Size-controlled quantum dots reveal the impact of intraband transitions on
  high-order harmonic generation in solids},}\ }\href@noop {} {\bibfield
  {journal} {\bibinfo  {journal} {Nature Physics}\ }\textbf {\bibinfo {volume}
  {18}},\ \bibinfo {pages} {874--878} (\bibinfo {year} {2022})}\BibitemShut
  {NoStop}%
\bibitem [{\citenamefont {Hirori}, \citenamefont {Sato},\ and\ \citenamefont
  {Kanemitsu}(2024)}]{hirori2024high}%
  \BibitemOpen
  \bibfield  {author} {\bibinfo {author} {\bibfnamefont {H.}~\bibnamefont
  {Hirori}}, \bibinfo {author} {\bibfnamefont {S.~A.}\ \bibnamefont {Sato}}, \
  and\ \bibinfo {author} {\bibfnamefont {Y.}~\bibnamefont {Kanemitsu}},\
  }\bibfield  {title} {\enquote {\bibinfo {title} {High-order harmonic
  generation in solids: The role of intraband transitions in extreme nonlinear
  optics},}\ }\href@noop {} {\bibfield  {journal} {\bibinfo  {journal} {The
  Journal of Physical Chemistry Letters}\ }\textbf {\bibinfo {volume} {15}},\
  \bibinfo {pages} {2184--2192} (\bibinfo {year} {2024})}\BibitemShut {NoStop}%
\bibitem [{\citenamefont {Leo}(1998)}]{leo1998interband}%
  \BibitemOpen
  \bibfield  {author} {\bibinfo {author} {\bibfnamefont {K.}~\bibnamefont
  {Leo}},\ }\bibfield  {title} {\enquote {\bibinfo {title} {Interband optical
  investigation of bloch oscillations in semiconductor superlattices},}\
  }\href@noop {} {\bibfield  {journal} {\bibinfo  {journal} {Semiconductor
  science and technology}\ }\textbf {\bibinfo {volume} {13}},\ \bibinfo {pages}
  {249} (\bibinfo {year} {1998})}\BibitemShut {NoStop}%
\bibitem [{\citenamefont {Ghimire}\ \emph {et~al.}(2011)\citenamefont
  {Ghimire}, \citenamefont {DiChiara}, \citenamefont {Sistrunk}, \citenamefont
  {Agostini}, \citenamefont {DiMauro},\ and\ \citenamefont
  {Reis}}]{ghimire2011observation}%
  \BibitemOpen
  \bibfield  {author} {\bibinfo {author} {\bibfnamefont {S.}~\bibnamefont
  {Ghimire}}, \bibinfo {author} {\bibfnamefont {A.~D.}\ \bibnamefont
  {DiChiara}}, \bibinfo {author} {\bibfnamefont {E.}~\bibnamefont {Sistrunk}},
  \bibinfo {author} {\bibfnamefont {P.}~\bibnamefont {Agostini}}, \bibinfo
  {author} {\bibfnamefont {L.~F.}\ \bibnamefont {DiMauro}}, \ and\ \bibinfo
  {author} {\bibfnamefont {D.~A.}\ \bibnamefont {Reis}},\ }\bibfield  {title}
  {\enquote {\bibinfo {title} {Observation of high-order harmonic generation in
  a bulk crystal},}\ }\href@noop {} {\bibfield  {journal} {\bibinfo  {journal}
  {Nature physics}\ }\textbf {\bibinfo {volume} {7}},\ \bibinfo {pages}
  {138--141} (\bibinfo {year} {2011})}\BibitemShut {NoStop}%
\bibitem [{\citenamefont {Shuai}\ \emph {et~al.}(2014)\citenamefont {Shuai},
  \citenamefont {Geng}, \citenamefont {Xu}, \citenamefont {Liao},\ and\
  \citenamefont {Andr{\'e}}}]{shuai2014charge}%
  \BibitemOpen
  \bibfield  {author} {\bibinfo {author} {\bibfnamefont {Z.}~\bibnamefont
  {Shuai}}, \bibinfo {author} {\bibfnamefont {H.}~\bibnamefont {Geng}},
  \bibinfo {author} {\bibfnamefont {W.}~\bibnamefont {Xu}}, \bibinfo {author}
  {\bibfnamefont {Y.}~\bibnamefont {Liao}}, \ and\ \bibinfo {author}
  {\bibfnamefont {J.-M.}\ \bibnamefont {Andr{\'e}}},\ }\bibfield  {title}
  {\enquote {\bibinfo {title} {From charge transport parameters to charge
  mobility in organic semiconductors through multiscale simulation},}\
  }\href@noop {} {\bibfield  {journal} {\bibinfo  {journal} {Chemical Society
  Reviews}\ }\textbf {\bibinfo {volume} {43}},\ \bibinfo {pages} {2662--2679}
  (\bibinfo {year} {2014})}\BibitemShut {NoStop}%
\bibitem [{\citenamefont {Jauho}\ and\ \citenamefont
  {Johnsen}(1996)}]{jauho1996dynamical}%
  \BibitemOpen
  \bibfield  {author} {\bibinfo {author} {\bibfnamefont {A.-P.}\ \bibnamefont
  {Jauho}}\ and\ \bibinfo {author} {\bibfnamefont {K.}~\bibnamefont
  {Johnsen}},\ }\bibfield  {title} {\enquote {\bibinfo {title} {Dynamical
  franz-keldysh effect},}\ }\href@noop {} {\bibfield  {journal} {\bibinfo
  {journal} {Physical review letters}\ }\textbf {\bibinfo {volume} {76}},\
  \bibinfo {pages} {4576} (\bibinfo {year} {1996})}\BibitemShut {NoStop}%
\bibitem [{\citenamefont {Nordstrom}\ \emph {et~al.}(1998)\citenamefont
  {Nordstrom}, \citenamefont {Johnsen}, \citenamefont {Allen}, \citenamefont
  {Jauho}, \citenamefont {Birnir}, \citenamefont {Kono}, \citenamefont {Noda},
  \citenamefont {Akiyama},\ and\ \citenamefont
  {Sakaki}}]{nordstrom1998excitonic}%
  \BibitemOpen
  \bibfield  {author} {\bibinfo {author} {\bibfnamefont {K.}~\bibnamefont
  {Nordstrom}}, \bibinfo {author} {\bibfnamefont {K.}~\bibnamefont {Johnsen}},
  \bibinfo {author} {\bibfnamefont {S.}~\bibnamefont {Allen}}, \bibinfo
  {author} {\bibfnamefont {A.-P.}\ \bibnamefont {Jauho}}, \bibinfo {author}
  {\bibfnamefont {B.}~\bibnamefont {Birnir}}, \bibinfo {author} {\bibfnamefont
  {J.}~\bibnamefont {Kono}}, \bibinfo {author} {\bibfnamefont {T.}~\bibnamefont
  {Noda}}, \bibinfo {author} {\bibfnamefont {H.}~\bibnamefont {Akiyama}}, \
  and\ \bibinfo {author} {\bibfnamefont {H.}~\bibnamefont {Sakaki}},\
  }\bibfield  {title} {\enquote {\bibinfo {title} {Excitonic dynamical
  franz-keldysh effect},}\ }\href@noop {} {\bibfield  {journal} {\bibinfo
  {journal} {Physical Review Letters}\ }\textbf {\bibinfo {volume} {81}},\
  \bibinfo {pages} {457} (\bibinfo {year} {1998})}\BibitemShut {NoStop}%
\bibitem [{\citenamefont {Potanina}\ \emph {et~al.}(2021)\citenamefont
  {Potanina}, \citenamefont {Flindt}, \citenamefont {Moskalets},\ and\
  \citenamefont {Brandner}}]{potanina2021thermodynamic}%
  \BibitemOpen
  \bibfield  {author} {\bibinfo {author} {\bibfnamefont {E.}~\bibnamefont
  {Potanina}}, \bibinfo {author} {\bibfnamefont {C.}~\bibnamefont {Flindt}},
  \bibinfo {author} {\bibfnamefont {M.}~\bibnamefont {Moskalets}}, \ and\
  \bibinfo {author} {\bibfnamefont {K.}~\bibnamefont {Brandner}},\ }\bibfield
  {title} {\enquote {\bibinfo {title} {Thermodynamic bounds on coherent
  transport in periodically driven conductors},}\ }\href@noop {} {\bibfield
  {journal} {\bibinfo  {journal} {Physical Review X}\ }\textbf {\bibinfo
  {volume} {11}},\ \bibinfo {pages} {021013} (\bibinfo {year}
  {2021})}\BibitemShut {NoStop}%
\bibitem [{\citenamefont {Luo}\ \emph {et~al.}(2019)\citenamefont {Luo},
  \citenamefont {Wang}, \citenamefont {Wang}, \citenamefont {Wang},
  \citenamefont {Sun},\ and\ \citenamefont {Liu}}]{luo2019high}%
  \BibitemOpen
  \bibfield  {author} {\bibinfo {author} {\bibfnamefont {H.}~\bibnamefont
  {Luo}}, \bibinfo {author} {\bibfnamefont {B.}~\bibnamefont {Wang}}, \bibinfo
  {author} {\bibfnamefont {E.}~\bibnamefont {Wang}}, \bibinfo {author}
  {\bibfnamefont {X.}~\bibnamefont {Wang}}, \bibinfo {author} {\bibfnamefont
  {Y.}~\bibnamefont {Sun}}, \ and\ \bibinfo {author} {\bibfnamefont
  {K.}~\bibnamefont {Liu}},\ }\bibfield  {title} {\enquote {\bibinfo {title}
  {High-responsivity photovoltaic photodetectors based on mote2/mose2 van der
  waals heterojunctions},}\ }\href@noop {} {\bibfield  {journal} {\bibinfo
  {journal} {Crystals}\ }\textbf {\bibinfo {volume} {9}},\ \bibinfo {pages}
  {315} (\bibinfo {year} {2019})}\BibitemShut {NoStop}%
\bibitem [{\citenamefont {Oka}\ and\ \citenamefont
  {Kitamura}(2019)}]{oka2019floquet}%
  \BibitemOpen
  \bibfield  {author} {\bibinfo {author} {\bibfnamefont {T.}~\bibnamefont
  {Oka}}\ and\ \bibinfo {author} {\bibfnamefont {S.}~\bibnamefont {Kitamura}},\
  }\bibfield  {title} {\enquote {\bibinfo {title} {Floquet engineering of
  quantum materials},}\ }\href@noop {} {\bibfield  {journal} {\bibinfo
  {journal} {Annual Review of Condensed Matter Physics}\ }\textbf {\bibinfo
  {volume} {10}},\ \bibinfo {pages} {387--408} (\bibinfo {year}
  {2019})}\BibitemShut {NoStop}%
\bibitem [{\citenamefont {Klausmeier}(2008)}]{klausmeier2008floquet}%
  \BibitemOpen
  \bibfield  {author} {\bibinfo {author} {\bibfnamefont {C.~A.}\ \bibnamefont
  {Klausmeier}},\ }\bibfield  {title} {\enquote {\bibinfo {title} {Floquet
  theory: a useful tool for understanding nonequilibrium dynamics},}\
  }\href@noop {} {\bibfield  {journal} {\bibinfo  {journal} {Theoretical
  Ecology}\ }\textbf {\bibinfo {volume} {1}},\ \bibinfo {pages} {153--161}
  (\bibinfo {year} {2008})}\BibitemShut {NoStop}%
\bibitem [{\citenamefont {Leskes}, \citenamefont {Madhu},\ and\ \citenamefont
  {Vega}(2010)}]{leskes2010floquet}%
  \BibitemOpen
  \bibfield  {author} {\bibinfo {author} {\bibfnamefont {M.}~\bibnamefont
  {Leskes}}, \bibinfo {author} {\bibfnamefont {P.}~\bibnamefont {Madhu}}, \
  and\ \bibinfo {author} {\bibfnamefont {S.}~\bibnamefont {Vega}},\ }\bibfield
  {title} {\enquote {\bibinfo {title} {Floquet theory in solid-state nuclear
  magnetic resonance},}\ }\href@noop {} {\bibfield  {journal} {\bibinfo
  {journal} {Progress in nuclear magnetic resonance spectroscopy}\ }\textbf
  {\bibinfo {volume} {57}},\ \bibinfo {pages} {345--380} (\bibinfo {year}
  {2010})}\BibitemShut {NoStop}%
\bibitem [{\citenamefont {Shu}\ \emph {et~al.}(2018)\citenamefont {Shu},
  \citenamefont {Liu}, \citenamefont {Cao}, \citenamefont {Yang}, \citenamefont
  {Zhang}, \citenamefont {Plenio}, \citenamefont {Jelezko},\ and\ \citenamefont
  {Cai}}]{shu2018observation}%
  \BibitemOpen
  \bibfield  {author} {\bibinfo {author} {\bibfnamefont {Z.}~\bibnamefont
  {Shu}}, \bibinfo {author} {\bibfnamefont {Y.}~\bibnamefont {Liu}}, \bibinfo
  {author} {\bibfnamefont {Q.}~\bibnamefont {Cao}}, \bibinfo {author}
  {\bibfnamefont {P.}~\bibnamefont {Yang}}, \bibinfo {author} {\bibfnamefont
  {S.}~\bibnamefont {Zhang}}, \bibinfo {author} {\bibfnamefont {M.~B.}\
  \bibnamefont {Plenio}}, \bibinfo {author} {\bibfnamefont {F.}~\bibnamefont
  {Jelezko}}, \ and\ \bibinfo {author} {\bibfnamefont {J.}~\bibnamefont
  {Cai}},\ }\bibfield  {title} {\enquote {\bibinfo {title} {Observation of
  floquet raman transition in a driven solid-state spin system},}\ }\href@noop
  {} {\bibfield  {journal} {\bibinfo  {journal} {Physical Review Letters}\
  }\textbf {\bibinfo {volume} {121}},\ \bibinfo {pages} {210501} (\bibinfo
  {year} {2018})}\BibitemShut {NoStop}%
\bibitem [{\citenamefont {Rechtsman}\ \emph {et~al.}(2013)\citenamefont
  {Rechtsman}, \citenamefont {Zeuner}, \citenamefont {Plotnik}, \citenamefont
  {Lumer}, \citenamefont {Podolsky}, \citenamefont {Dreisow}, \citenamefont
  {Nolte}, \citenamefont {Segev},\ and\ \citenamefont
  {Szameit}}]{rechtsman2013photonic}%
  \BibitemOpen
  \bibfield  {author} {\bibinfo {author} {\bibfnamefont {M.~C.}\ \bibnamefont
  {Rechtsman}}, \bibinfo {author} {\bibfnamefont {J.~M.}\ \bibnamefont
  {Zeuner}}, \bibinfo {author} {\bibfnamefont {Y.}~\bibnamefont {Plotnik}},
  \bibinfo {author} {\bibfnamefont {Y.}~\bibnamefont {Lumer}}, \bibinfo
  {author} {\bibfnamefont {D.}~\bibnamefont {Podolsky}}, \bibinfo {author}
  {\bibfnamefont {F.}~\bibnamefont {Dreisow}}, \bibinfo {author} {\bibfnamefont
  {S.}~\bibnamefont {Nolte}}, \bibinfo {author} {\bibfnamefont
  {M.}~\bibnamefont {Segev}}, \ and\ \bibinfo {author} {\bibfnamefont
  {A.}~\bibnamefont {Szameit}},\ }\bibfield  {title} {\enquote {\bibinfo
  {title} {Photonic floquet topological insulators},}\ }\href@noop {}
  {\bibfield  {journal} {\bibinfo  {journal} {Nature}\ }\textbf {\bibinfo
  {volume} {496}},\ \bibinfo {pages} {196--200} (\bibinfo {year}
  {2013})}\BibitemShut {NoStop}%
\bibitem [{\citenamefont {Tiwari}, \citenamefont {Gu},\ and\ \citenamefont
  {Franco}(2023)}]{tiwari2023floquet}%
  \BibitemOpen
  \bibfield  {author} {\bibinfo {author} {\bibfnamefont {V.}~\bibnamefont
  {Tiwari}}, \bibinfo {author} {\bibfnamefont {B.}~\bibnamefont {Gu}}, \ and\
  \bibinfo {author} {\bibfnamefont {I.}~\bibnamefont {Franco}},\ }\bibfield
  {title} {\enquote {\bibinfo {title} {Floquet theory and computational method
  for the optical absorption of laser-dressed solids},}\ }\href@noop {}
  {\bibfield  {journal} {\bibinfo  {journal} {Physical Review B}\ }\textbf
  {\bibinfo {volume} {108}},\ \bibinfo {pages} {064308} (\bibinfo {year}
  {2023})}\BibitemShut {NoStop}%
\bibitem [{\citenamefont {Zheng}\ and\ \citenamefont
  {Wang}(2023)}]{zheng2023multiple}%
  \BibitemOpen
  \bibfield  {author} {\bibinfo {author} {\bibfnamefont {F.}~\bibnamefont
  {Zheng}}\ and\ \bibinfo {author} {\bibfnamefont {L.-w.}\ \bibnamefont
  {Wang}},\ }\bibfield  {title} {\enquote {\bibinfo {title} {Multiple k-point
  nonadiabatic molecular dynamics for ultrafast excitations in periodic
  systems: The example of photoexcited silicon},}\ }\href@noop {} {\bibfield
  {journal} {\bibinfo  {journal} {Physical Review Letters}\ }\textbf {\bibinfo
  {volume} {131}},\ \bibinfo {pages} {156302} (\bibinfo {year}
  {2023})}\BibitemShut {NoStop}%
\bibitem [{\citenamefont {Xie}\ \emph {et~al.}(2022)\citenamefont {Xie},
  \citenamefont {Xu}, \citenamefont {Wang},\ and\ \citenamefont
  {Zhuang}}]{xie2022surface}%
  \BibitemOpen
  \bibfield  {author} {\bibinfo {author} {\bibfnamefont {H.}~\bibnamefont
  {Xie}}, \bibinfo {author} {\bibfnamefont {X.}~\bibnamefont {Xu}}, \bibinfo
  {author} {\bibfnamefont {L.}~\bibnamefont {Wang}}, \ and\ \bibinfo {author}
  {\bibfnamefont {W.}~\bibnamefont {Zhuang}},\ }\bibfield  {title} {\enquote
  {\bibinfo {title} {Surface hopping dynamics in periodic solid-state materials
  with a linear vibronic coupling model},}\ }\href@noop {} {\bibfield
  {journal} {\bibinfo  {journal} {The Journal of Chemical Physics}\ }\textbf
  {\bibinfo {volume} {156}} (\bibinfo {year} {2022})}\BibitemShut {NoStop}%
\bibitem [{\citenamefont {Krotz}\ and\ \citenamefont
  {Tempelaar}(2022)}]{krotz2022reciprocal}%
  \BibitemOpen
  \bibfield  {author} {\bibinfo {author} {\bibfnamefont {A.}~\bibnamefont
  {Krotz}}\ and\ \bibinfo {author} {\bibfnamefont {R.}~\bibnamefont
  {Tempelaar}},\ }\bibfield  {title} {\enquote {\bibinfo {title} {A
  reciprocal-space formulation of surface hopping},}\ }\href@noop {} {\bibfield
   {journal} {\bibinfo  {journal} {The Journal of Chemical Physics}\ }\textbf
  {\bibinfo {volume} {156}} (\bibinfo {year} {2022})}\BibitemShut {NoStop}%
\bibitem [{\citenamefont {Chen}, \citenamefont {Wang},\ and\ \citenamefont
  {Dou}(2024)}]{chen2024floquet}%
  \BibitemOpen
  \bibfield  {author} {\bibinfo {author} {\bibfnamefont {J.}~\bibnamefont
  {Chen}}, \bibinfo {author} {\bibfnamefont {Y.}~\bibnamefont {Wang}}, \ and\
  \bibinfo {author} {\bibfnamefont {W.}~\bibnamefont {Dou}},\ }\bibfield
  {title} {\enquote {\bibinfo {title} {Floquet nonadiabatic mixed
  quantum--classical dynamics in periodically driven solid systems},}\
  }\href@noop {} {\bibfield  {journal} {\bibinfo  {journal} {The Journal of
  Chemical Physics}\ }\textbf {\bibinfo {volume} {160}} (\bibinfo {year}
  {2024})}\BibitemShut {NoStop}%
\bibitem [{\citenamefont {Krotz}, \citenamefont {Provazza},\ and\ \citenamefont
  {Tempelaar}(2021)}]{krotz2021reciprocal}%
  \BibitemOpen
  \bibfield  {author} {\bibinfo {author} {\bibfnamefont {A.}~\bibnamefont
  {Krotz}}, \bibinfo {author} {\bibfnamefont {J.}~\bibnamefont {Provazza}}, \
  and\ \bibinfo {author} {\bibfnamefont {R.}~\bibnamefont {Tempelaar}},\
  }\bibfield  {title} {\enquote {\bibinfo {title} {A reciprocal-space
  formulation of mixed quantum--classical dynamics},}\ }\href@noop {}
  {\bibfield  {journal} {\bibinfo  {journal} {The Journal of Chemical Physics}\
  }\textbf {\bibinfo {volume} {154}} (\bibinfo {year} {2021})}\BibitemShut
  {NoStop}%
\bibitem [{\citenamefont {Mosallanejad}, \citenamefont {Chen},\ and\
  \citenamefont {Dou}(2023)}]{mosallanejad2023floquet}%
  \BibitemOpen
  \bibfield  {author} {\bibinfo {author} {\bibfnamefont {V.}~\bibnamefont
  {Mosallanejad}}, \bibinfo {author} {\bibfnamefont {J.}~\bibnamefont {Chen}},
  \ and\ \bibinfo {author} {\bibfnamefont {W.}~\bibnamefont {Dou}},\ }\bibfield
   {title} {\enquote {\bibinfo {title} {Floquet-driven frictional effects},}\
  }\href@noop {} {\bibfield  {journal} {\bibinfo  {journal} {Physical Review
  B}\ }\textbf {\bibinfo {volume} {107}},\ \bibinfo {pages} {184314} (\bibinfo
  {year} {2023})}\BibitemShut {NoStop}%
\bibitem [{\citenamefont {Wang}\ and\ \citenamefont
  {Dou}(2023)}]{wang2023nonadiabatic}%
  \BibitemOpen
  \bibfield  {author} {\bibinfo {author} {\bibfnamefont {Y.}~\bibnamefont
  {Wang}}\ and\ \bibinfo {author} {\bibfnamefont {W.}~\bibnamefont {Dou}},\
  }\bibfield  {title} {\enquote {\bibinfo {title} {Nonadiabatic dynamics near
  metal surface with periodic drivings: A floquet surface hopping algorithm},}\
  }\href@noop {} {\bibfield  {journal} {\bibinfo  {journal} {The Journal of
  Chemical Physics}\ }\textbf {\bibinfo {volume} {158}} (\bibinfo {year}
  {2023})}\BibitemShut {NoStop}%
\bibitem [{\citenamefont {Wang}\ \emph {et~al.}(2024)\citenamefont {Wang},
  \citenamefont {Mosallanejad}, \citenamefont {Liu},\ and\ \citenamefont
  {Dou}}]{wang2024nonadiabatic}%
  \BibitemOpen
  \bibfield  {author} {\bibinfo {author} {\bibfnamefont {Y.}~\bibnamefont
  {Wang}}, \bibinfo {author} {\bibfnamefont {V.}~\bibnamefont {Mosallanejad}},
  \bibinfo {author} {\bibfnamefont {W.}~\bibnamefont {Liu}}, \ and\ \bibinfo
  {author} {\bibfnamefont {W.}~\bibnamefont {Dou}},\ }\bibfield  {title}
  {\enquote {\bibinfo {title} {Nonadiabatic dynamics near metal surfaces with
  periodic drivings: A generalized surface hopping in floquet
  representation},}\ }\href@noop {} {\bibfield  {journal} {\bibinfo  {journal}
  {Journal of Chemical Theory and Computation}\ }\textbf {\bibinfo {volume}
  {20}},\ \bibinfo {pages} {644--650} (\bibinfo {year} {2024})}\BibitemShut
  {NoStop}%
\end{thebibliography}%

\end{document}